\documentstyle[12pt]{article}

\newcommand{\mrm}[1]{\mathrm{#1}}
\newmathalphabet*{\mbf}{cmr}{b}{n}
\newmathalphabet*{\mii}{cmr}{m}{it}
\newmathalphabet*{\mtt}{cmtt}{m}{n}

\newcommand{\mW}{m_{\mathrm W}}
\newcommand{\GW}{\Gamma_{\mathrm W}}
\newcommand{\overmW}{\overline{m}_{\mathrm W}}

\newcommand{\pT}{p_{\perp}}

\newcommand{\boldp}{\mbf{p}}
\newcommand{\boldx}{\mbf{x}}
\newcommand{\boldu}{\mbf{u}}
\newcommand{\boldv}{\mbf{v}}
\newcommand{\boldbeta}{\mbox{\boldmath $\beta$}}
\newcommand{\boldzero}{\mbf{0}}

\newcommand{\Py}{\mbox{\sc Pythia}}
\newcommand{\Je}{\mbox{\sc Jetset}}

\renewcommand{\b}{{\mathrm b}}
\renewcommand{\c}{{\mathrm c}}
\renewcommand{\d}{{\mathrm d}}
\newcommand{\e}{{\mathrm e}}

\newcommand{\g}{{\mathrm g}}

\newcommand{\p}{{\mathrm p}}
\newcommand{\q}{{\mathrm q}}
\newcommand{\s}{{\mathrm s}}
\renewcommand{\t}{{\mathrm t}}
\renewcommand{\u}{{\mathrm u}}

\newcommand{\B}{{\mathrm B}}
\newcommand{\D}{{\mathrm D}}

\renewcommand{\H}{{\mathrm H}}
\newcommand{\J}{{\mathrm J}}

\newcommand{\W}{{\mathrm W}}
\newcommand{\Z}{{\mathrm Z}}
\newcommand{\bbar}{\overline{\mathrm b}}
\newcommand{\cbar}{\overline{\mathrm c}}
\newcommand{\dbar}{\overline{\mathrm d}}

\newcommand{\pbar}{\overline{\mathrm p}}
\newcommand{\qbar}{\overline{\mathrm q}}
\newcommand{\sbar}{\overline{\mathrm s}}
\newcommand{\tbar}{\overline{\mathrm t}}
\newcommand{\ubar}{\overline{\mathrm u}}

\newcommand{\JP}{\J/\psi}

\newcommand{\ee}{\e^+\e^-}

\newcommand{\mmax}{\mrm{max}}

\newenvironment{Itemize}{\begin{list}{$\bullet$}%
{\setlength{\topsep}{0.2mm}\setlength{\partopsep}{0.2mm}%
\setlength{\itemsep}{0.2mm}\setlength{\parsep}{0.2mm}}}%
{\end{list}}
\newcounter{enumct}
\newenvironment{Enumerate}{\begin{list}{\arabic{enumct}.}%
{\usecounter{enumct}\setlength{\topsep}{0.2mm}%
\setlength{\partopsep}{0.2mm}\setlength{\itemsep}{0.2mm}%
\setlength{\parsep}{0.2mm}}}{\end{list}}
\newlength{\captivewidth}
\newcommand{\captive}[1]{\rule{5mm}{0mm}%
\begin{minipage}{\captivewidth}%
\caption[small]{#1}\end{minipage}}

\begin{document}

\sloppy

\pagestyle{empty}

\begin{flushright}
CERN-TH.7011/93 \\
DTP/93/74
\end{flushright}

\vspace{\fill}

\begin{center}
{\LARGE\bf On Colour Rearrangement}\\[4mm]
{\LARGE\bf in Hadronic W$^+$W$^-$ Events}\\[10mm]
{\Large Torbj\"orn Sj\"ostrand} \\[3mm]
{\large Theory Division, CERN} \\[1mm]
{\large CH-1211 Geneva 23, Switzerland}\\[5mm]
{\large and} \\[5mm]
{\Large Valery A. Khoze} \\[3mm]
{\large Department of Physics, University of Durham} \\[1mm]
{\large Durham DH1 3LE, England} \\
\end{center}

\vspace{\fill}

\begin{center}
\bf{Abstract}
\end{center}
\vspace{-0.5\baselineskip}
\noindent
We discuss the possibility of colour rearrangement in
$\ee \to \W^+ \W^- \to \q_1 \qbar_2 \q_3 \qbar_4$ events,
i.e.\ that the original colour singlets $\q_1 \qbar_2$ and
$\q_3 \qbar_4$ may be transmuted, for instance, into new
singlets $\q_1 \qbar_4$ and $\q_3 \qbar_2$. The
effects on event properties could be quite large if such a
rearrangement would occur instantaneously, so that
gluon emission would be restricted to each of the new singlets
separately. We argue that such a scenario is unlikely for two
reasons. Firstly, the $\W^+$ and $\W^-$ usually decay at
separate times after the $\W^+\W^-$ production, which leads to
large relative phases for energetic radiation off the two
constituents of a rearranged system, and a corresponding dampening
of the QCD cascades.
Secondly, within the perturbative scenario the colour transmutation
appears only in order $\alpha_s^2$ and is colour-suppressed.
Colour reconnection at longer time scales is quite feasible,
however, and may affect the fragmentation phase.
If so, the nature of non-perturbative QCD can be probed in a
new way. We formulate several alternative toy models and use
these to estimate the colour reconnection probability as a
function of the event kinematics. Possible consequences for LEP~2
events are illustrated, with special attention to systematic
errors in $\W$ mass determinations.

\vspace{\fill}

\noindent
CERN-TH.7011/93 \\
October 1993

\clearpage

\pagestyle{plain}
\setcounter{page}{1}

\section{Introduction}

Consider the production of several colour singlet particles, which
decay to coloured partons close to each other in space and time.
A topical example, which will be used as a basis for the continued
discussion, is
\begin{equation}
\e^+ \e^- \to \W^+ \W^- \to \q_1 \qbar_2 \, \q_3 \qbar_4
\label{process}
\end{equation}
at LEP 2. The outgoing partons first undergo a perturbative phase,
in which showers of additional partons develop, and subsequently
a non-perturbative phase, in which the partons fragment into
a hadronic final state. In this paper we will study the extent to
which the showering and fragmentation of one singlet can `interfere'
with that of the other, in such a way that observable event
properties are affected. This is a topic about which little is
known today, but which has the potential to provide new insights,
as we shall try to show.

Since perturbative QCD is reasonably well understood, it is
possible to predict with some confidence what to expect on the
partonic level. We will here demonstrate that within the purely
perturbative scenario these interference effects should be
negligibly small. However, non-perturbative QCD is not yet well
understood, and there is no obvious reason why interference effects
should be small in the fragmentation process. It is therefore in
this area that process (\ref{process}) can be a very useful probe.
We will develop and compare the predictions of two main alternative
models, which correspond to two different hypotheses on the
structure of the QCD vacuum and of the
confinement mechanism. With a lot of hard work and some luck,
LEP 2 could be in a position to provide some discrimination.

To illustrate what we mean by `interference', consider process
(\ref{process}) above. The standard point of
view is to assume the $\W^+$ and $\W^-$ decays to be independent
of each other. Apart from some spin-related correlations in the
angular distribution of the $\W$-decay products, each $\W$ system
can then shower and fragment without any reference to what is
happening to the other. In particular, the perturbative parton emission
is initiated by two separate colour dipoles, $\q_1 \qbar_2$
and $\q_3 \qbar_4$. This picture should be valid if the
$\W$'s are long-lived, so that the two $\W$ decays occur at
well-separated points. In the other extreme, the $\W$'s are assumed
to decay `instantaneously', so that the process really involves the
production of a $\q_1 \qbar_2 \q_3 \qbar_4$ state in a single point.
This state forms one single colour quadrupole, and it is the quadrupole
as a whole that can emit additional partons and eventually fragment
to hadrons. No trace is left of the individual identities of the
$\W^+$ and $\W^-$ decay products.

It is very difficult to obtain a complete description along the
latter lines --- all the standard methods of dealing with the
showering and fragmentation processes would be of little use.
(The matrix element calculation approach would still be valid, but
more difficult to apply.) However, our experience with perturbative
QCD in $\q\qbar\g$ events has taught us that a colour quadrupole
can be well approximated by the sum of two separate colour dipoles,
each of which is a net colour singlet. (There may also be
contributions from non-singlet dipoles, see below.) The new aspect
is that we here have two sets of such potential dipoles, the original
one, i.e.\ $\q_1 \qbar_2$ and $\q_3 \qbar_4$, and a colour rearranged
set, $\q_1 \qbar_4$ and $\q_3 \qbar_2$.

The possibility of having events with rearranged colour dipoles
was first studied by Gustafson, Pettersson and Zerwas (GPZ)
\cite{GPZ}. Since they assume that
the dipole mass sets the scale for the amount of
energetic gluon radiation and multiple soft particle production to
be expected, event properties may change a lot if the original dipoles
are replaced by the rearranged ones. The original dipoles each have
the $\W$ mass, while the rearranged ones may have much lower masses,
if the kinematics is selected suitably. GPZ could therefore present
large differences in the total multiplicity, the energy flow, the
rapidity distribution, and so on.

Unfortunately, this is not a very likely scenario. As we will show,
the separation between the $\W^+$ and $\W^-$ decay vertices is large
enough that energetic QCD radiation will occur independently within
each original dipole, to a good first approximation. However, the
fragmentation process is extending much further in space and time,
and thus colour reconnection in this phase is a possibility.
Observable event properties are less affected by a reconnection
that comes only after the perturbative phase \cite{ee500}, but this
does not necessarily mean that effects are negligibly small. Detailed
studies of event shapes should help reveal the nature and size of
colour rearrangement phenomena.

While interference effects are interesting in their own right, if
there, they may also provide a serious source of uncertainty.
One of the main objectives of LEP 2 will be to determine the W mass.
This could either be done by a threshold scan of the cross section,
or by a reconstruction of W masses event by event for some fixed energy
above the $\W^+\W^-$ threshold. Currently the latter approach is the
favoured one \cite{LEP2work}. Fully hadronic decays, i.e.\ process
(\ref{process}), would seem to be preferable: jet energies need not
be so well measured in the calorimeters, since already a knowledge
of the four jet directions would be enough to constrain the
kinematics and therefore the W masses. Electroweak effects, including
initial state radiation, and effects of semileptonic $\c$ decays
would be under control. By contrast, purely leptonic or mixed
leptonic--hadronic decays suffer from a large missing momentum
vector given by the neutrino(s).

The main problem is that the assignment of hadrons to either of the
two $\W$'s is not unique. If only an experimental issue, the smearing
could be estimated and corrected for. However, to this we now add
an uncertainty in the colour structure of the event. So long as the
original $\q_1 \qbar_2$ and $\q_3 \qbar_4$ dipoles shower and
fragment independently of each other, the invariant masses of the
hadrons belonging to these two systems, if separated correctly, add
up to give the original $\W^+$ and $\W^-$ masses, respectively. But
if a reconnection occurs, e.g. to the alternative colour singlets
$\q_1 \qbar_4$ and $\q_3 \qbar_2$ (or even more complicated systems),
there is no concept of a conserved
$\W$ mass in the fragmentation. This introduces an additional element
of smearing and, more dangerously, the potentiality for a systematic
bias in the $\W$ mass determination. Clearly, such effects must be
studied and understood, in order to be corrected for.

The possibility of colour rearrangements is not restricted to
$\ee \to \W^+\W^-$ events. Other examples could have been found:
$\p\p/\p\pbar \to \W^+ \W^-$, $\ee \to \Z^0 \Z^0$, $\ee \to \Z^0 \H^0$,
$\p\p/\p\pbar \to \W^{\pm}\H^0$, $\H^0 \to \W^+ \W^-$,
$\t \tbar \to \b \W^+ \bbar \W^-$, etc. One could also add processes
that involve a single colour singlet particle interfering with the
beam jets of a $\p\p$ event, such as $\p\p \to \W^{\pm}$, or even
interactions among beam jets themselves, such as in multiple
parton--parton interaction events or in heavy ion collisions.
The problem with the latter processes is that there are so many other
uncertainties that any chance of systematic studies would be excluded.
By contrast, in $\ee \to \W^+\W^-$ events, the decay of a
non-interfering $\W$ pair is very well understood, thanks to our
experience with $\Z^0$ decays at LEP 1.

The hadronic decay of a $\B$ meson is another typical example of a
colour quadrupole structure. $\JP$ production in such decays is the
one place where colour reconnection has already been observed (see
discussion and additional references in ref. \cite{GPZ}): in a
\mbox{$\b \to \c + \W^- \to \c + \cbar + \s$} decay the $\c$ and
$\cbar$ belong to separate colour singlets, but appear almost
simultaneously and can therefore coherently form a colour singlet
state such as the $\JP$. Of course, the number of colours not being
infinite, there is a finite chance of $1/N_C^2 = 1/9$ that the
$\c$ and $\cbar$ are in a colour singlet configuration. However,
this kind of accidental colour ambiguities is not what we imply by
colour rearrangement --- after all, had the $\W$ been a long-lived
particle, so that the $\c$ and $\cbar$ were produced far away,
the question of whether they are in a relative octet or singlet
state would have been moot. (Accidental ambiguities do not spoil
the original colour singlet combinations, such as $\cbar + \s$,
but only add more.) Further, while difficult to estimate, it seems
that the colour reconnection probability in $\B$ decay is more
like 20\%, i.e.\ higher than the above maximum. In this paper we will
therefore concentrate on dynamical models for colour rearrangement,
and leave aside the accidental possibility. Anyway, one should not
use the details of $\B \to \JP$ phenomenology to provide any specific
input for $\W^+ \W^-$ production, since the space--time evolution
and the capability to radiate energetic gluons are
rather different in the two processes.

To summarize, there are two main reasons to study the phenomenon
of colour rearrangement:
\begin{Enumerate}
\item
In its own right, since it provides a laboratory for a better
understanding of the space--time structure of perturbative and
non-perturbative QCD.
\item
As a potential source of error to $\W$ mass determinations and other
measurements. (In the end, the precision of the theoretical
predictions has to match the experimental accuracy, or even exceed
it.)
\end{Enumerate}
We cannot today predict what will come out of experimental
studies at LEP 2, but want to provide here some comments on what
one could plausibly expect under some simplified alternative
scenarios. This might form a reasonable starting point for more
refined theoretical and experimental studies in the future.

The plan of the paper is therefore as follows. In section 2 we
show what maximal effects could be expected if colour rearrangement
always took place, either `instantaneously' or only later,
between the parton shower and the fragmentation phases. These
`worst case' or `best case' scenarios, depending on the point of view,
are subsequently rejected in favour of more realistic ones. In
Section 3 we show why one would not expect sizeable effects on the
perturbative level, based on the time separation between
the $\W^+$ and $\W^-$ decays, and on colour algebra. The
subsequent discussion is therefore
concentrated on effects in the fragmentation process: in Section
4 we describe different ways of estimating the colour rearrangement
probability as a function of event topology, and in Section 5 some
experimental consequences are discussed. Finally, Section 6 contains
a summary and outlook.

\section{Maximal effects of colour reconnection}

The objective of this section is purely didactic: to familiarize
readers with possible experimental consequences of colour
rearrangements. Our `best bet' estimates of such effects will be
presented in Section 5. The current section is here mainly because
it is needed to appreciate the significance of Sections 3 and 4.
The na\"{\i}ve scenarios we introduce here should be taken for what
they are: extremely simple recipes for colour reconnection, in which
a number of known complications are swept under the carpet. In
particular, the $\W^+$ and $\W^-$ are assumed to decay
instantaneously, so that there is no space or time separation
between the two decay vertices.

In order to stay as close as possible to the studies of GPZ, we
first consider two $\W$ bosons produced at rest, each with the
nominal mass $\mW = 80$ GeV, decaying into the final state
$\u\dbar + \d\ubar$, with an opening angle of $30^{\circ}$
between $\u$ and $\ubar$, Fig. 1a. To first approximation,
one may then entertain three alternatives:
\begin{Enumerate}
\item In the no-reconnection scenario, the $\u\dbar$ and $\d\ubar$
systems shower and fragment independently of each other, Fig. 1b.
The event therefore looks like two overlayed $\ee$ annihilation
events at rest, each with 80 GeV c.m. energy. This scenario is
just what one expects from an extreme perturbative point of view.
\item In the GPZ `instantaneous' reconnection scenario, two colour
singlets $\u\ubar$ and $\d\dbar$ are immediately formed, each then
having an invariant mass of 20.7 GeV and a net motion away from the
origin. These systems subsequently shower and fragment independently
of each other, Fig. 1c. There is only little radiation, since the
maximum scale for that radiation is 20 GeV rather than 80 GeV.
The final event looks like two strongly boosted
20 GeV $\ee$ annihilation events. We do not know of any physics
mechanism that would lead to this scenario (except maybe as a
rare fluctuation), but include it since it
was advocated by GPZ and since it gives maximal effects.
\item In the intermediate reconnection scenario, a reconnection occurs
between the shower and fragmentation steps. Therefore the original
$\u\dbar$ and $\d\ubar$ colour singlet systems shower, each with
maximum virtuality 80 GeV, just as in the no-reconnection scenario.
Stricty speaking, the gluons emitted in the $\u\dbar$ system are not
uniquely associated to either the $\u$ or the $\dbar$. However, in
practice, the shower algorithm used \cite{MBTS} does allow a pragmatic
subdivision of the radiation into two subsets, essentially (but not
quite) by hemisphere. For instance, if partons are ordered in
colour/string/dipole order, a $\W^+$ ($\W^-$) branching to
$\u\g_1^+\g_2^+\g_3^+\dbar$ ($\d\g_1^-\g_2^-\g_3^-\g_4^-\ubar$)
may be split into two subsets $\u\g_1^+$ and $\g_2^+\g_3^+\dbar$
($\d\g_1^-\g_2^-\g_3^-$ and $\g_4^-\ubar$). After the showers, the
partons can therefore be reconnected into two new colour singlets,
$\u\g_1^+\g_4^-\ubar$ and $\d\g_1^-\g_2^-\g_3^-\g_2^+\g_3^+\dbar$.
Each of these systems then fragments separately, Fig. 1d. In the limit
$\alpha_s \to 0$, i.e.\ when energetic QCD radiation is switched off,
this coincides with the instantaneous scenario. In practice, however,
QCD radiation is profuse, and the masses of the two fragmenting systems
is substantially above the 20 GeV nominal value.
\end{Enumerate}
Fragmentation is assumed to be given according to the Lund string model
\cite{Lund} (see also Section 4.1). Events are generated with the
{\Je} 7.3 program \cite{Jetset}, which is known to well describe
the $\ee \to \q\qbar$ events in the 10--100 GeV mass range.

Some simple results are presented in Fig. 2. The event axis
(with respect to which e.g.\ rapidity is defined) is here
the theoretical one, i.e.\ intermediate between the original
$\u$ and $\ubar$ directions. The total multiplicity
is about a factor of 2 lower for the instantaneous reconnection
scenario: $\langle n_{\mrm{ch}} \rangle = 21.8$ versus 40.0 for no
reconnection and 37.8 for the intermediate one. The first number
reflects the lower invariant mass of the colour singlet systems.
The instantaneous
scenario contains two strongly boosted systems with nothing in
between; therefore the rapidity distribution has a very strong dip
in the middle, Fig. 2b, which is reflected in the multiplicity
distribution for the central rapidity region, $|y| < 1$, Fig. 2c.
It is also visible in the charged particle flow in the event plane,
Fig. 2d, where particle production at around $90^{\circ}$ to the
event axis is suppressed. It is maybe a bit more surprising that
the instantaneous scenario does not have more particle production
than the no-reconnection one at $0^{\circ}$; after all, this is the
region where a string connects the $\u$ ($\d$) and $\ubar$ ($\dbar$)
partons of the instantaneous events, while no such string is
present in the no-reconnection ones. Our interpretation is that this
shows how large the effects of QCD radiation are, i.e.\ that radiation
with an 80 GeV maximum virtuality scale leads to an overwhelmingly
large broadening of jet profiles.

In all the distributions, the intermediate scenario deviates much less
from the standard one than does the instantaneous. This means that
the particle flow is to a large extent dominated by the perturbative
gluon emission phase, at least when this phase is assumed to be valid
down to a cut-off scale of $Q_0 \approx 1$ GeV and gluon emission
is therefore profuse. However, the effects of the intermediate
scenario seem puny only when compared with the instantaneous one: the
differences between the no- and intermediate-reconnection
distributions of Fig. 2b (e.g.) would be easily distinguishable
experimentally.

The problem is that the kinematics we have considered so far is
totally unrealistic. While effects are largest when the two $\W$'s
are produced at rest, this is the point where the phase space for
production is vanishing. Next we therefore have to introduce some
semi-realistic experimental conditions.

In order to generate complete
$\ee \to \W^+ \W^- \to \q_1 \qbar_2 \q_3 \qbar_4$ events, we use the
{\Py}~5.6 event generator \cite{Pythia}. The $\W^+\W^-$ pair is here
distributed according to the product of three terms: a Breit-Wigner
for each $\W$, and a basic $\ee \to \W^+\W^-$ matrix element (which is
a function of the actual $\W^{\pm}$ masses) \cite{Brown}. Subsequently
each $\W$ decays to a $\q\qbar$ pair, $\u\dbar$, $\u\sbar$, $\c\dbar$,
$\c\sbar$ or their charge conjugates. The decay angles are properly
correlated \cite{Gunion}. This is followed by perturbative parton
shower evolution and non-perturbative fragmentation, as above.
Initial state photon radiation can and should also be included for
detailed experimental studies, but is omitted for most of the studies
below.

To illustrate the size of the effects, consider a c.m. energy of
170 GeV, i.e.\ some distance above the threshold. The nominal input
$\W$ mass is 80 GeV; the average generated $\mW$ is somewhat lower,
predominantly because of phase-space effects.The charged
multiplicity distribution is shown in Fig. 3a for the scenarios
1--3 above. Clearly differences are nowhere as drastic as in
Fig. 2a: the average values are 36.6 without recoupling, 36.2 with
intermediate, and 33.7 with instantaneous recoupling. In part this
is related to an averaging over various event topologies in Fig. 3,
where Fig. 2 was only for one particularly favourable configuration.
In part one also expects smaller differences further away above
threshold, since the difference in mass between original and
reconnected colour systems is reduced. The latter effect is
particularly easy to understand if one considers the region far above
threshold, where the two $\W$'s and their respective decay products
are strongly boosted away from each other, and a reconnection between
the two widely separated systems in fact leads to an increase in
system masses, i.e.\ opposite to the threshold behaviour.

The charged rapidity distribution with respect to the thrust axis,
Fig. 3b, and the number of charged particles in $|y|<1$, Fig. 3c,
also show much lesser differences than those observed in
Figs. 2b and 2c. In particular, the change in kinematics leads to
a much narrower central rapidity dip. Qualitatively, however,
differences are still there. This is particularly obvious in the
low-multiplicity part of Fig. 3c. Differences can be enhanced by
various cuts, e.g.\ by requiring a large thrust value.

As a final exercise, we study the task of $\W$ mass determination.
This is an important topic in itself \cite{LEP2work}, and it is not
our intention here to optimize an algorithm so as to minimize
statistical errors. Rather, the objective is to find whether any
systematic effects arise from reconnections.

The details of our $\W$ mass reconstruction algorithm will be
presented in Section 5.3. For the moment, suffice it to say that
four jets are reconstructed per event (events without a clear
four-jet structure are rejected), assuming that particle
four-momenta are fully known. The jets are paired to give two
$\W$ masses, which are then averaged to give one
$\overmW = (m_{\W^+} + m_{\W^-})/2$
number per event. The difference between the reconstructed and the
generated $\overmW$ is shown in Fig. 3d. By the procedure
adopted, the distributions thus do not contain any spread from the
intrinsic Breit-Wigner shape of the $\W$'s or from
detector imperfections. Any spread comes from misassignments of
particles to jets. Large deviations may occur when entire subjets
are incorrectly clustered.

If one considers the range $\pm 10$ GeV of difference between
reconstructed and generated masses, the average and spread is
$-0.28 \pm 1.57$ GeV for the no-reconnection scenario,
$-0.14 \pm 1.60$ GeV for the intermediate one, and $0.30 \pm 1.55$ GeV
for the instantaneous one. The fraction of $\W$'s with deviations
larger than $\pm10$ GeV is 1.5\%, 1.5\% and 1.7\%, respectively.
Compared with the no-reconnection scenario, the intermediate
(instantaneous) one gives a systematic shift of over 100 (500)
MeV. The aimed-for statistical error on the $\W$ mass is roughly
50 MeV. It is therefore of importance to understand whether these
colour reconnection numbers above have to be included as a
systematical uncertainty, or whether reconnection effects are only
a small fraction of this.

\section{The perturbative picture of colour rearrangement}

In this section we want to discuss the colour dynamics of particle
flow in $\W^+\W^-$ events from a purely perturbative point of view.
The non-perturbative standpoint will follow in Section 4.

\subsection{Introduction to the perturbative approach}

During the last years experiments, especially at the $\Z^0$ pole,
have provided an exceedingly rich source of information on the
jet structure of final states in hard processes (see e.g.\
ref. \cite{K1}). These data have shown that, at high energies,
the main characteristics of multihadronic events are determined
by the perturbative stage of the process evolution.
Analytical perturbation theory --- the perturbative approach
\cite{K2} --- provides a quantitative description of
inclusive particle production in jets. In the perturbative
approach the resummation of
the perturbative series is performed, and the terms of relative order
$\sqrt{\alpha_s}$ are taken into account in a systematic way ---
the modified leading logarithmic approximation (MLLA). Under the
key assumption that the non-perturbative hadronization process is
local in the configuration space of partons, the infrared
singularities can be factorized out of hadron distributions.
If so, the asymptotic shapes of these distributions are fully
predicted --- the hypothesis of local parton--hadron duality (LPHD).
Data agree remarkably well with the predictions of the MLLA--LPHD
framework \cite{K3}.

Until now, the perturbative approach was applied only to systems of
partons produced almost simultaneously, with a short time scale
\begin{equation}
t_{\mrm{prod}} \sim \frac{1}{E} \ll \frac{1}{\mu} ~,
\label{tprodmu}
\end{equation}
where $E$ is the hard scale of the production process and
$\mu^{-1} \approx 1$ fm is the characteristic strong interaction
time scale. The radiation accompanying such a system
can be represented as a superposition of gauge
invariant terms, in which each external quark line is uniquely
connected to an external antiquark line of the same colour.
The description of gluons is straightforward: remember that a gluon
has both a quark and an antiquark colour index. The system is thus
decomposed into a set of colourless $\q\qbar$ antennae (dipoles).
One of the simplest examples is the celebrated $\q\qbar\g$ system,
which (to leading order in $1/N_C^2$) is well approximated by the
sum of two separate antennae/dipoles. The perturbative treatment of
the $\q\qbar$ antennae is based on the following key ideas:
\begin{Enumerate}
\item
The principal source of multiple hadroproduction is gluon
bremsstrahlung caused by conserved colour currents. Therefore the
flow of colour quantum numbers, reflecting the dynamics at short
distances, controls the particle distributions in the final state.
\item
The evolution of a quark jet is
viewed as a sequence of coherent parton branchings.
In the final state, an original quark jet is enshrouded by
secondary partons resulting from radiation of quasi-collinear
and/or soft gluons with momenta $k$ and transverse momenta
$k_{\perp}$ in the range
\begin{equation}
\mu \lesssim k_{\perp} \ll k \ll E ~.
\end{equation}
It is the large probability for such emissions,
\begin{equation}
{\cal W} \sim \frac{\alpha_s}{\pi} \, \ln^2 E \sim 1
\end{equation}
that leads to the well-known double logarithmic phenomena in jet
development.
\end{Enumerate}
In this approach, the particle multiplicity is obtained
as the convolution of the probability of primary gluon bremstrahlung
off the parent quarks with the multiplicity initiated by such a
gluon \cite{K2}.

Any restriction on the primary gluon energy to be well below $E$
\begin{equation}
k \lesssim k_{\mmax} \ll E
\end{equation}
drastically reduces the perturbatively induced multiplicity and
makes the $\q\qbar$ antenna practically inactive \cite{K2,K4}.
As we shall see, this is of relevance for colour rearranged systems
in $\W^+\W^-$ events.

\subsection{W-pair decays in the perturbative approach}

Encouraged by the successes of the perturbative approach, one could
be tempted to apply the same quark--gluon dynamics to the description
of the final state in process (\ref{process}). In particular, each
individual $\W^{\pm}$ decay corresponds to a parton shower at almost
the same scale as the well-studied $\Z^0 \to \q\qbar$ case.

As we shall show, the emission of a single primary gluon (which
subsequently initiates a coherent parton shower) corresponds to
the no-reconnection scenario. Within the perturbative approach,
colour transmutations can result only from the interferences between
gluons (virtual as well as real) radiated in the two decays.
Rearrangements of the colour flows should lead to a dependence of
the structure of final particle distributions on the relative angles
between the jets originating from the different $\W$'s (on top of what
may be there from trivial kinematics).

It is the goal of this section to demonstrate why the colour
reconnection effects, viewed perturbatively, are strongly suppressed.
Our argumentation below is based on two main reasons, which are
deeply rooted in the basic structure of QCD:
\begin{Enumerate}
\item
Because of the group structure of QCD, at least two gluons should be
emitted to generate the colour rearrangement. Moreover the interference
piece proves to be suppressed by $1/N_C^2$ as compared to the
${\cal O}(\alpha_s^2)$ no-reconnection emissions.
\item
The effects of the $\W$ width $\GW$ strongly restrict the energy
range of primary gluons generated by the alternative systems
of type $\q_1\qbar_4$ and $\q_3\qbar_2$. Not so far from the
$\W^+\W^-$ threshold one expects
\begin{equation}
k \lesssim k_{\mmax}^{\mrm{recon}} \sim \GW ~.
\end{equation}
Therefore the would-be
parton showers initiated by such systems are terminated at a
virtuality scale of ${\cal O}(\GW)$, and can hardly lead to
sizeable fluctuations in the structure of the hadronic final state.
\end{Enumerate}

\subsection{Single-gluon emission in W pair decays}

Let us first consider the emission of a single primary soft gluon
of four-momentum $k$, see Fig. 4. The momenta of the final state
quarks are labelled by
$\ee \to \q_1(p_1) \, \qbar_2(p_2) \, \q_3(p_3) \, \qbar_4(p_4)$,
with $Q_1 = p_1 + p_2$ and $Q_2 = p_3 + p_4$. Denote by
$\widehat{M}^{(0)}$ the non-radiative Feynman amplitude with the
$\W$ propagators removed and introduce the conserved currents
$J^{\mu}(k)$, $J'^{\mu}(k)$ generated by the individual $\W$ decays.
In the limit $k \ll p_i$ the amplitude $M^{(1)}$ for gluon
radiation accompanying process (\ref{process}) is then
\begin{equation}
M^{(1)} \equiv (M^{(1)})_{jm}^{in} = g_s \, \widehat{M}^{(0)} \,
\left[ (T^a)_j^i \, \delta_m^n \, (J(k)\cdot\epsilon_{\lambda}) +
\delta_j^i \, (T^a)_m^n \, (J'(k)\cdot\epsilon_{\lambda}) \right] ~,
\end{equation}
where $g_s$ is related to the strong coupling constant by
$\alpha_s = g_s^2/4\pi$, $\epsilon_{\lambda}$ is the gluon
polarization vector, $T^a$ are the {\bf SU(3)} colour matrices,
and $a = 1,\ldots,8$ and $i,j,m,n = 1,2,3$ are colour labels.

The $\W^{\pm}$ propagator functions $D$
are absorbed into the definition of the currents $J^{\mu}(k)$,
$J'^{\mu}(k)$ \cite{K5}
\begin{eqnarray}
J^{\mu}(k) & = & j^{\mu}(k) \, D(Q_1+k) \, D(Q_2) ~,
\nonumber \\
J'^{\mu}(k) & = & j'^{\mu}(k) \, D(Q_1) \, D(Q_2+k) ~,
\end{eqnarray}
where
\begin{eqnarray}
j^{\mu}(k) & = & \frac{p_1^{\mu}}{p_1 \cdot k} -
\frac{p_2^{\mu}}{p_2 \cdot k} ~,
\nonumber \\
j'^{\mu}(k) & = & \frac{p_3^{\mu}}{p_3 \cdot k} -
\frac{p_4^{\mu}}{p_4 \cdot k}
\end{eqnarray}
describe the gauge-invariant soft gluon emission from two colour
charges of momenta $p_1,p_2$ ($p_3,p_4$). The expression for the
propagator function is
\begin{equation}
D(Q) = \frac{1}{Q^2 - \mW^2 + i \mW\GW} ~.
\end{equation}
To emphasize that the emission is generated primarily by the
conserved currents $j^{\mu}$, $j'^{\mu}$ we have introduced the
`radiative blobs' in Fig. 4 and in what follows.

In order to obtain the differential distribution for gluon emission
we have to square the matrix element, sum over colours and spins,
and integrate over the $Q_1^2$ and $Q_2^2$ virtualities.
We find (see also \cite{K4,K5})
\begin{eqnarray}
\frac{1}{\sigma^{(0)}} \, \d\sigma^{(1)} & = &
\frac{\d^3k}{\omega} \, \frac{C_F \, \alpha_s}{4\pi^2} \,
{\cal F}^{(1)} ~, \\
{\cal F}^{(1)} & = & \left( \frac{\mW\GW}{\pi} \right)^2
\int \d Q_1^2 \, \d Q_2^2 \,
\left[ - J \cdot J^{\dagger} - J' \cdot J'^{\dagger} \right]
{}~,
\end{eqnarray}
where $\omega = k^0$ is the gluon energy and
$C_F = (N_C^2-1)/2N_C = 4/3$.

In the massless quark limit the radiation pattern ${\cal F}^{(1)}$
is given by
\begin{equation}
{\cal F}^{(1)} = 2 \, (\widehat{12} + \widehat{34} ) ~,
\end{equation}
where the $\q\qbar$ `antennae' are defined by \cite{K2}
\begin{equation}
\widehat{ij} = \frac{(p_i \cdot p_j)}{(p_i \cdot k)(p_j \cdot k)} ~.
\end{equation}
Near threshold, where the decay products of a $\W$ are almost
back-to-back,
\begin{eqnarray}
{\cal F}^{(1)} & = & \frac{2}{\omega^2} \, \left(
\frac{1-\cos\theta_{12}}{(1-\cos\theta_1)(1-\cos\theta_2)} +
\frac{1-\cos\theta_{34}}{(1-\cos\theta_3)(1-\cos\theta_4)} \right)
\nonumber \\
& \approx & \frac{4}{\omega^2} \, \left(
\frac{1}{\sin^2\theta_1} + \frac{1}{\sin^2\theta_3} \right) ~,
\label{Foneth}
\end{eqnarray}
where $\theta_i$ is the angle between parton $i$ and the gluon,
and $\theta_{ij}$ the angle between partons $i$ and $j$.

Let us emphasize that on the level of single gluon emission, real as
well as virtual, the two antennae $\q_1\qbar_2$ and $\q_3\qbar_4$
do not interact and the colour flows are not rearranged. The absence
of such an interaction is easily seen from the diagram of Fig. 5,
which represents the decay--decay radiative interference contribution
to the cross section of process (\ref{process}). At the position of the
vertical dashed line, both the $\q_1 \qbar_2$ and the $\q_3\qbar_4$
subsystems have to be in colour singlet states (in order to couple to
the $\W$'s), so the gluon octet charge is uncompensated.

\subsection{Double-gluon interference effects in W-pair decays}

At least two primary gluons, real or virtual, should be emitted to
generate a colour flow rearrangement, see Figs. 6--8. Note that the
diagrams of Figs. 6a and 6b do not interfere with each other, and that
the diagrams of Fig. 7 could interfere with those of Fig. 4, thus
inducing a colour transmutation. The infrared divergences in the
virtual pieces are cancelled by the corresponding real emissions.
For the case of decay--decay radiative interference the soft
emissions are cancelled in the inclusive cross section up to at least
${\cal O}(\GW/\mW)$ (see \cite{K6,K7} and below).

The main qualitative results for the reconnection effects appearing in
${\cal O}(\alpha_s^2)$ are not much different for various decay--decay
interference samples. We shall examine below one example corresponding
to the diagrams of Fig. 6a. In the limit $k_1,k_2 \ll p_i$ the matrix
element can be written as
\begin{eqnarray}
M_a^{(2)} = g_s^2 \, \widehat{M}^{(0)} & & \left[ (T^a)_j^i \,
(T^b)_m^n \, (j(k_1)\cdot\epsilon_{\lambda}^{(1)}) \,
(j'(k_2)\cdot\epsilon_{\lambda'}^{(2)}) \,
D(Q_1 + k_1) \, D(Q_2 + k_2) \right. \nonumber \\
& & \left. + \left\{ a \leftrightarrow b,
k_1 \leftrightarrow k_2, \epsilon_{\lambda}^{(1)} \leftrightarrow
\epsilon_{\lambda'}^{(2)} \right\} \right] ~,
\end{eqnarray}
where $k_{1,2}$ are the momenta of the soft gluons and
$\epsilon_{\lambda}^{(1)}$, $\epsilon_{\lambda'}^{(2)}$ are their
polarization vectors.

After summing over colours and spins, the interference term may be
presented in the form
\begin{eqnarray}
\frac{1}{\sigma_{0}} \, \d\sigma_a^{\mrm{int}} & \simeq &
\frac{\d^3 k_1}{\omega_1} \, \frac{\d^3 k_2}{\omega_2} \,
\left( \frac{C_F \, \alpha_s}{4\pi^2} \right)^2 \,
\frac{1}{N_C^2 -1} \, \frac{1}{2} \, {\cal F}_a^{\mrm{int}} ~,
\label{sigaint} \\
 {\cal F}_a^{\mrm{int}} & = & 2 \, \chi_{12} \,
(j(k_1) \cdot j'(k_1))
\, (j(k_2) \cdot j'(k_2)) ~,
\end{eqnarray}
with
\begin{equation}
- (j(k) \cdot j'(k)) = \widehat{14} + \widehat{23} -
\widehat{13} - \widehat{24} ~.
\label{Fintdipol}
\end{equation}
Here $\chi_{12}$ is the so-called profile function \cite{K4,K5},
which controls the decay--decay interferences:
\begin{equation}
\chi_{12} = \left( \frac{\mW \GW}{\pi} \right)^2
\; \Re \int \d Q_1^2 \, \d Q_2^2 \, D(Q_1 + k_1) \, D^*(Q_1 + k_2)
\, D(Q_2 + k_2) \, D^*(Q_2 + k_1) ~,
\label{proffun}
\end{equation}
where $D^*$ is the complex conjugate of $D$ and $\Re$ represents
the real part. The profile
function has the formal property that $\chi_{12} \to 0$ as
$\GW \to 0$ and $\chi_{12} \to 1$ as $\Gamma \to \infty$.

The interference is suppressed by $1/(N_C^2-1) = 1/8$ as compared
to the total rate of double primary gluon emissions (related to
the square of diagrams of the type of Fig. 6), see eq.
(\ref{sigaint}). This is a result of the ratio of the corresponding
colour traces,
\begin{equation}
\frac{ \mrm{Tr}(T^a T^b) \cdot \mrm{Tr}(T^a T^b) }
{ \mrm{Tr}(T^a T^a) \cdot \mrm{Tr}(T^b T^b) } =
\frac{ (C_F \, N_C)/2}{(C_F \, N_C)^2} = \frac{1}{N_C^2 - 1}.
\end{equation}
Such a suppression takes place for any decay--decay radiative
interference piece, real as well as virtual, as is clear from
Fig. 8.

Near threshold and in the limit of massless quarks the interference
contribution to the radiation pattern is
\begin{equation}
{\cal F}_a^{\mrm{int}} = \frac{2 \, \chi_{12}}{\omega_1^2 \,
\omega_2^2} \,
\; \frac{16 \, \cos\phi_{13} \, \cos\tilde{\phi}_{13}}
{\sin\theta_1 \, \sin\theta_3 \, \sin\tilde{\theta}_1 \,
\sin\tilde{\theta}_3} ~,
\label{Faintang}
\end{equation}
where $\theta_i$ ($\tilde{\theta}_i$) is the angle between the $\q_i$
and the gluon $k_1$ ($k_2$), and $\phi_{13}$ ($\tilde{\phi}_{13}$)
is the relative azimuth between $\q_1$ and $\q_3$ around the direction
of the $k_1$ ($k_2$). The expression in eq. (\ref{Faintang})
evidently contains a dependence on the relative orientation of the
decay products of the two $\W$'s. (The interference is maximal when
all the partons lie in the same plane,
$\phi_{13} = \tilde{\phi}_{13} = 0$, cf. eqs. (\ref{Foneth}) and
(\ref{Faintang}).) Therefore one might expect that the decay--decay
interferences would induce some colour-suppressed reconnection
effects in the structure of final states in process (\ref{process}).

\subsection{W width effects}

It is the profile function $\chi_{12}$ that cuts down the phase space
available for gluon emissions by the alternative quark pairs (or by any
accidental colour singlets) and thus eliminates the very possibility
for the reconnected systems to develop QCD cascades. That the $\W$
width does control the radiative interferences can be easily
understood by considering the extreme cases.

If the $\W$-boson lifetime could be considered as very short,
$1/\GW \to 0$, both the $\q_1\qbar_2$ and $\q_3\qbar_4$
pairs appear almost instantaneously, and they radiate coherently,
as though produced at the same vertex. In the other
extreme, $\GW \to 0$, the $\q_1\qbar_2$ and $\q_3\qbar_4$
pairs appear at very different times $t_1$, $t_2$ after the
$\W^+\W^-$ production,
\begin{equation}
t_{\mrm{prod}} \sim \frac{1}{\mW} \ll \Delta t = |t_1 - t_2| \sim
\frac{1}{\GW} ~.
\end{equation}
The two dipoles therefore radiate gluons and produce hadrons according
to the no-reconnection scenario.

The crucial point is the proper choice of the scale the $\W$ width
should be compared with. That scale is set by the energies of primary
emissions, real or virtual \cite{K4,K5,K6}. Let us clarify this
supposing, for simplicity, that we are in the $\W^+\W^-$ threshold
region. The relative phases of radiation accompanying two $\W$ decays
are then given by the quantity
\begin{equation}
\omega_i \, \Delta t \sim \frac{\omega_i}{\GW} ~.
\end{equation}
When $\omega_i/\GW \gg 1$ the phases fluctuate wildly and
the interference terms vanish. This is a direct consequence of the
radiophysics of the colour flows \cite{K2} reflecting the wave
dynamics of QCD. The argumentation remains valid for energies above
the $\W^+\W^-$ threshold as well. Suppression of the interference
in the case of radiation with $\omega_i \gg \GW$ can be
demonstrated also in a more formal way.

One can perform the integration over $\d Q_1^2$ and $\d Q_2^2$
in eq. (\ref{proffun}) by taking the residues of the poles in the
propagators. This gives
\begin{equation}
\chi_{12} = \frac{ \mW^2 \GW^2 \left( \kappa_1 \kappa_2
+ \mW^2 \GW^2 \right) }{ \left( \kappa_1^2 +
\mW^2 \GW^2 \right) \left( \kappa_2^2 +
\mW^2 \GW^2 \right) } ~,
\label{chigen}
\end{equation}
with
\begin{equation}
\kappa_{1,2} = Q_{1,2} \cdot (k_1 - k_2) ~.
\end{equation}
For the interference between the diagrams of Fig. 6b, the
corresponding profile function is given by the same formula
with $k_2 \to - k_2$. Near the $\W^+\W^-$ pair threshold
eq. (\ref{chigen}) is reduced to
\begin{equation}
\chi_{12} = \frac{\GW^2}{\GW^2 +
(\omega_1 - \omega_2)^2} ~.
\label{chithresh}
\end{equation}

 From eq. (\ref{chithresh}) it is clear that only primary emissions
with $\omega_{1,2} \lesssim \GW$ can induce significant
rearrangement effects: the radiation of energetic gluons (real or
virtual) with $\omega_{1,2} \gg \GW$ pushes the $\W$
propagators far off their non-radiative resonant positions, so that
the propagator functions $D(Q_1 + k_1)$ and $D(Q_1 + k_2)$
($D(Q_2 + k_1)$ and $D(Q_2 + k_2)$) corresponding to the same $\W$
practically do not overlap. We can neglect the
contribution to the inclusive cross section from kinematical
configurations with $\omega_1, \omega_2 \gg \GW$,
\mbox{$|\omega_1 - \omega_2| \lesssim \GW$} since the
corresponding phase-space volume is negligibly small.

Equation (\ref{chigen}) clearly shows that $\chi_{12}$
vanishes if any of the scalar products $Q_i \cdot k_j$ ($i,j = 1,2$)
well exceeds $\mW\GW$. Again accidental kinematics with
$\kappa_1, \kappa_2 \ll \mW\GW$ is suppressed because of
phase space reasons. Hence all our arguments concerning cutting
down the QCD cascades induced by the alternative systems remain valid
above the threshold as well. The smallness of the decay--decay radiative
interference for energetic emission in the production of a heavy
unstable particle pair, at the threshold and far above it, proves
to be of a general nature. For the case of $\ee \to \b\W^+\bbar\W^-$
this was explicitly demonstrated in ref. \cite{K5}.

At very high energies $E \gg \mW$ the energy scale of the would-be
QCD showers generated by the alternative systems is restricted more
strongly,
\begin{equation}
k_{\mmax}^{\mrm{recon}} \sim \frac{\GW \, \mW}{E_{\W}} ~.
\end{equation}
Within the framework of a perturbative analysis such a restriction
makes sense only if
\begin{equation}
\eta =  \mu \, \frac{E_{\W}}{\mW \, \GW} \lesssim 1 ~.
\end{equation}
Remembering that the lifetime of a $\W$ in the laboratory frame is
\begin{equation}
t_{\mrm{dec}} \sim \frac{E_{\W}}{\mW} \, \frac{1}{\GW}
\end{equation}
one can easily see that as long as $\eta < 1$, the requirement of
perturbative soft gluons $\omega > \mu$ automatically implies that a
$\W$ decays before the formation of the first light hadrons from the
QCD cascades.

In the extreme ultrarelativistic limit, when
$\eta = \mu \, t_{\mrm{dec}} \gg 1$, the energy of primary perturbative
gluons becomes limited from below, $\omega > \mu \, \eta$.
This restriction arises from the relationship between
$t_{\mrm{dec}}$ and the gluon hadronization time,
\begin{equation}
t_{\mrm{dec}} < t_{\mrm{had}} \sim \frac{\omega}{\mu^2} ~.
\end{equation}
The profile function $\chi_{12}$ in such an extreme case decreases
with increasing energy and the decay--decay radiative interference
is strongly suppressed. For instance,
\begin{eqnarray}
\chi_{12} \sim \left( \frac{\mu}{\omega} \right)^2 \,
\frac{1}{\eta^2} < \frac{1}{\eta^4} & ~~~~ & \mrm{for}~~
\mu \, \eta < \omega < \GW ~,      \nonumber \\
\chi_{12} < \frac{\mW^2}{E_{\W}^2} & ~~~~ & \mrm{for}~~
\omega > \GW ~.
\end{eqnarray}

Therefore, in addition to the $\alpha_s^2$ and $1/N_C^2$ suppression
effects noted above, any reconnected quark system (including
accidental colour singlets) proves to be practically inactive.
The bulk of radiation (and thus of multihadron production) in the
final state of process (\ref{process}) is governed by the original
$\q_1\qbar_2$ and $\q_3\qbar_4$ antennae, which radiate the
primary gluons with $\omega \gg \GW$ that initiate coherent
showers. The corresponding hard scale for the non-reconnected parton
showers is $\mW$. Also accounting for cascade multiplication, the
yield of the reconnection-sensitive particles can be quantified as
the multiplicity at a hard scale of ${\cal O}(\GW)$.

All the argumentation based on the effects of a phase difference
between the radiations accompanying two $\W$ decays remain valid
also in the case when one of the $\W$ bosons is practically real and
the other is far off the mass shell, i.e.\ $t_1 \sim 1/\GW$
and $t_2 \sim 1/\mW$. This case could be of interest for the
intermediate mass Higgs decay \cite{ee500}. The effects of the
profile function $\chi_{12}$ are the same for
$t_2 \ll t_1 \sim 1/\GW$ as when
$t_1,t_2 \sim 1/\GW$.

\subsection{Structure of inclusive particle flow in W-pair events}

Let us discuss the general topology of events corresponding to
process (\ref{process}). The single-inclusive particle flow
(antenna--dipole pattern) may be written as (see ref. \cite{K2}
for details)
\begin{equation}
\frac{8\pi \, \d N}{\d \Omega_{\mbf{n}}} =
2 \, \left[ (12) + (34) \right] \,
N'_{\q}\left( \frac{\mW}{2}\right)
+ R \, \left[ (14) + (23) - (13) - (24) \right] \,
N'_{\q}\left( k_{\mmax}^{\mrm{recon}}\right) ~.
\label{nflow}
\end{equation}
The distribution $(ij)$ describes the angular radiation
pattern of the $\widehat{ij}$ antenna,
\begin{equation}
(ij) \equiv \omega^2 \, (\widehat{ij}) =
\frac{a_{ij}}{a_i \, a_j} = \frac{1 - \mbf{n}_i \mbf{n}_j}%
{(1 - \mbf{n}_i \mbf{n}) (1 - \mbf{n}_j \mbf{n})} ~,
\end{equation}
where the $\mbf{n}_{i,j}$ denote the directions of the
$\q/\qbar$ momenta and $\mbf{n}$ the direction of the registered flow.
The factors
\begin{equation}
N'_{\q}( E_{\mrm{jet}}) = \frac{\d}{\d \, \ln E_{\mrm{jet}}}
N_{\q}(E_{\mrm{jet}})~,
\end{equation}
with $N_{\q}(E_{\mrm{jet}})$ the multiplicity inside a QCD jet of
energy/hardness $E_{\mrm{jet}}$, take into account the cascade particle
multiplication. Approximately \cite{K2},
\begin{equation}
\frac{N'_{\q}}{N_{\q}} \simeq \sqrt{\frac{2 N_C \alpha_s}{\pi}} \,
(1 + {\cal O}(\sqrt{\alpha_s})) ~.
\end{equation}

The factor $R$ describes the rearrangement strength. In principle,
it could be computed within the perturbative scenario. However,
for the purposes of this paper we shall here present only some
order-of-magnitude estimates. Each squared diagram of Figs. 4,
6 and 7 (as well as the set of corresponding higher-order diagrams)
contribute to the no-recoupling first term in eq. (\ref{nflow}).
Interferences between the diagrams of Fig. 6a and Fig. 6b and
between the diagrams of Fig. 7 and Fig. 4 exemplify the
lowest-order contribution to the reconnecting interference piece.

It follows from the discussion in the previous subsections that the
rearrangement coefficient is expected to be
\begin{equation}
R \lesssim {\cal O} \left( \frac{\alpha_s}{N_C^2} \right) ~.
\label{Rsuppr}
\end{equation}
Moreover, the suppression of energetic radiation accompanying the
reconnected systems makes the corresponding cascades practically
sterile, so the rearrangement affects only a few particles,
\begin{equation}
\frac{N_{\q}(k_{\mmax}^{\mrm{recon}})}{N_{\q}(\mW/2)} \sim
{\cal O}(10^{-1}) ~.
\end{equation}
Because of both these factors, the magnitude of the reconnection
effects in the perturbative scenario is expected to be numerically
small, ${\cal O}(10^{-2})$ or less. This gives the factor by which
the maximal perturbative effects shown in Section 2 should be
scaled down for a realistic estimate of perturbative reconnection
effects.

One can derive the antenna pattern corresponding to colour
transmutation (the second term in eq. \ref{nflow}), for instance
by examining the interference between the diagrams of Fig.~7 with
those of Fig. 4. The same structure appears for the interference
contributions corresponding to the diagrams of Fig. 6a and Fig. 6b
after integration over the momentum of one of the emitted gluons.
The infrared divergences corresponding to the unobserved gluon are
cancelled when both real and virtual emission contributions are
taken into account.

 From the antenna patterns given by eqs. (\ref{Fintdipol}) and
(\ref{nflow}) one immediately sees that, in addition to the two
rearranged dipoles $\q_1\qbar_4$ and $\q_3\qbar_2$ present in the
GPZ string picture, two other terms $\q_1\q_3$ and $\qbar_2\qbar_4$
appear. As was mentioned before, these terms are intimately
connected with the conservation of colour currents. Moreover, the
$\widehat{13}$ and $\widehat{24}$ antennae come in with a negative
sign. In general, QCD radiophysics predicts both attractive and
repulsive forces between quarks and antiquarks, see refs.
\cite{K2,K8,K9}. Normally the repulsion effects are quite small,
but in the case of colour-suppressed phenomena they may play an
important r\^ole.

One can easily understand the physical origin of the attraction and
repulsion effects with the help of the `QED' model of ref.
\cite{K9}, where quarks are replaced by leptons. For illustration,
the photonic interference pattern in
\begin{equation}
\gamma\gamma \to \Z^0\Z^0 \to \e^+\e^-\mu^+\mu^-
\label{QEDtoy}
\end{equation}
could be examined. In addition to the attractive forces between
opposite electrical charges ($\widehat{\e^-\mu^+}$ and
$\widehat{\e^+\mu^-}$ QED-antennae) there is a negative-sign
contribution ($\widehat{\e^-\mu^-}$ and $\widehat{\e^+\mu^+}$
QED-antennae) corresponding to the repulsive forces between two
same-sign charges. In QED there is no equivalent to the colour
suppression factor, so in the limit $\Gamma_{\Z} \to \infty$
(i.e. $\chi_{12} \to 1$) the dipole radiation structure is simply
given by the expression
\begin{equation}
\widehat{\e^-\e^+} + \widehat{\mu^-\mu^+} +
\left( \widehat{\e^-\mu^+} + \widehat{\e^+\mu^-}
- \widehat{\e^-\mu^-} - \widehat{\e^+\mu^+} \right) ~.
\end{equation}
For instance, near the $\Z^0\Z^0$ threshold, the total interference
is maximal and constructive (destructive) when $\e^-$ is
collinear with $\mu^-$ ($\mu^+$); for $\Gamma_{\Z} \to \infty$
the radiation pattern is equivalent to that induced by a
charge $-2$ (charge 0) particle.

By contrast to the perturbative QCD description, only the
positive-sign dipoles appear in the Lund string model. This is
because each string corresponds to a colour singlet, while the
negative-sign $\q\q$/$\qbar\qbar$ dipoles correspond to
non-singlets (antitriplets/sextets).
There need not be a physics conflict between the two pictures:
one should remember that the perturbative approach describes
short-distance phenomena, where partons may be considered free
to first approximation, while the Lund string picture is a model
for the long-distance behaviour of QCD, where confinement effects
should lead to a subdivision of the full system into colour singlet
subsystems (ultimately hadrons) with screened interactions between
these subsystems.

Also the r\^ole of colour quantum numbers may differ, as follows.
Reconnection is suppressed by a factor $1/(N_C^2-1)$ in the
perturbative description. The same suppression would appear in
the non-perturbative string model if only endpoint quark colours
were considered. For instance, in the
$\W^+\W^- \to \q_1 \qbar_2 \q_3 \qbar_4$ process, with perturbative
gluon emission neglected for the moment, the $\q_1\qbar_4$ system
would have to be in a colour singlet state for a reconnection to
be possible. However, in a realistic picture, the string itself
is made up of a multitude (an infinity) of coloured confinement
gluons. Therefore all colours are well represented in the local
neighbourhood of any potential string reconnection point. The local
gluons can then always be rearranged in such a way that colour
neutrality is maintained for the reconnected systems. Although some
suppression of the reconnection probability might well remain,
as a first guess we will assume that there is no such suppression
in the non-perturbative phase.

Let us come now to the issue of observability of the reconnection
effects in a real-life experiment. Analogously to the string
\cite{Lundstr} / drag \cite{K8} effect, colour rearrangement could
generate azimuthal anisotropies in the distributions of the particle
flow. That is, in the rest frame of one $\W$ the particle distribution
relative to the daughter-quark direction could become azimuthally
asymmetric (on top of the trivial kinematical effects caused by the
overlap with the decay products of the other $\W$). Such an asymmetry
should be strongly dependent on the overall topology of the
4-jet $\q_1\qbar_2\q_3\qbar_4$ system. It is instructive to note that
the negative-sign (13) and (24) dipoles of eq. (\ref{nflow})
in fact act to increase the magnitude of the string-like anisotropy
effects of the (14) and (23) dipoles.

It should be emphasized that, analogously to the other
colour-suppressed interference phenomena (see refs. \cite{K2,K8,K10}),
the rearrangement phenomenon can be viewed only on a completely
inclusive basis, when all the antennae--dipoles are simultaneously
active in the particle production. The very fact that the reconnection
pieces are not positive-definite reflects their wave interference
nature. Therefore the effects of recoupled sterile cascades should
appear on top of a background generated by the ordinary-looking
no-reconnection dipoles.

Again there is an important difference between the perturbative QCD
radiophysics picture and the non-perturbative string model. The latter
not only allows but even requires a completely exclusive description:
in the end the $\q_1\qbar_2\q_3\qbar_4$ system must be subdivided into
and fragment as two separate colour singlets, either
$\q_1\qbar_2$ and $\q_3\qbar_4$ or $\q_1\qbar_4$ and $\q_3\qbar_2$.
(Neglecting the fact that the recoupling and fragmentation will
involve additional partons emitted in the preceding perturbative
phase, see below.) The string model therefore predicts effects that
should be searched for on an event-by-event basis. Normally (such as
in the $\ee\to\q\qbar\g$ process) the two pictures work in quite
peaceful coexistence; differences only become drastic when dealing with
the small colour-suppressed effects.

Summing up the above discussion, it can be concluded that colour
rearrangement affects only a few low energy particles. Not so far
from the $\W^+\W^-$ threshold the magnitude of the
reconnection-induced anisotropy effects in the particle-flow
distribution is expected to be
\begin{equation}
\frac{\Delta N^{\mrm{recon}}}{N^{\mrm{no-recon}}} \lesssim
\frac{\alpha_s(\GW)}{N_C^2} \,
\frac{N'_{\q}(k_{\mmax}^{\mrm{recon}})}{N'_{\q}(\mW/2)}
\lesssim {\cal O}(10^{-2}) ~.
\end{equation}

In the integral inclusive cross section for
$\ee \to \W^+\W^- \to \q_1\qbar_2\q_3\qbar_4$, at and above the
$\W^+\W^-$ threshold, the reconnection effects are expected to
be negligibly small,
\begin{equation}
\frac{\Delta\sigma^{\mrm{recon}}}{\sigma} \lesssim
\frac{(C_F \, \alpha_s)^2}{N_C^2} \, \frac{\GW}{\mW} ~,
\label{fracrecon}
\end{equation}
where we would expect that the running coupling constant should be
evaluated at a scale of ${\cal O}(\GW)$. Numerically, then,
$\sigma^{\mrm{recon}}/\sigma \ll 10^{-3}$.

Let us clarify the origin of the factor $\GW/\mW$ in eq.
(\ref{fracrecon}) (for details see ref. \cite{K6}).
Because of the exact cancellation between the real and virtual
soft ($|k^0| \ll \mW$) gluon emissions, the interference
rearrangement effects can manifest themselves only in terms
of the order of $|k^0|/\mW$. But the radiated energy in the
reconnected systems is restricted to be in the
$|k^0| \lesssim \GW$ domain, and so the magnitude of the
rearrangement phenomena should include the factor
$\GW/\mW$. Note that the soft gluon cancellation argument is
based on an integration over all gluon momenta; it does not
apply for the registered particle flow, which is a more
exclusive distribution.

Using the analogy with the QED radiation in reaction (\ref{QEDtoy}),
the anisotropy of particle flow in QCD can be put in correspondence
with that of photon emission in (\ref{QEDtoy}). The rearrangement
effects in the integrated QCD process cross section would correspond
to the interference radiative corrections to the cross section of
process (\ref{QEDtoy}). The essential difference between the
radiative phenomena in the two processes (\ref{process}) and
(\ref{QEDtoy}) is that in the QED case the interference terms
contribute to the photon angular distribution (for
$\omega \lesssim \Gamma_{\Z}$) with the same strength as the
independent emission terms of each $\Z^0$ decay. Only in the QCD
case does the decay--decay recoupling interference acquire a small
weighting factor, see eq. (\ref{Rsuppr}).

\section{Non-perturbative models for topology dependence}

Having demonstrated that perturbative colour rearrangement effects
are negligibly small,
in the rest of the paper we consider exclusively the possibility of
reconnection occurring as a part of the non-perturbative fragmentation
phase. Since fragmentation is not understood from first principles,
this requires model building rather than exact calculations. We will
use the standard Lund string fragmentation model \cite{Lund}
as a starting point,
as summarized in Section 4.1. The colour reconnection phenomenon
is therefore equated with the possibility that the string drawing
given by the preceding hard process and parton shower activity
is subsequently modified.
We expect the reconnection probability to depend on the detailed
string topology, i.e.\ to vary as a function of c.m. energy, actual
$\W$ masses, the amount of parton shower activity and the angles
between outgoing partons.

Throughout this section, the discussion is entirely on the
probabilistic level, i.e. any negative-sign interference effects
are absent. This means that the original colour singlets
$\q_1\qbar_2$ and $\q_3\qbar_4$ may transmute to new singlets
$\q_1\qbar_4$ and $\q_3\qbar_2$, but that any effects e.g.\ of the
$\q_1\q_3$ and $\qbar_2\qbar_4$ dipoles (cf. eq. (\ref{Fintdipol}))
are absent. In this respect, the non-perturbative discussion is
more limited in outlook than the perturbative one above.

The imagined time sequence is the following (for details see
Section 4.2). The $\W^+$ and $\W^-$ fly apart from their common
production vertex and decay at some distance. Around each of these
decay vertices, a perturbative parton shower evolves from an original
$\q\qbar$ pair. The typical distance that a virtual parton (of mass
$m \sim 10$~GeV) travels
before branching is comparable with the typical $\W^+\W^-$
separation, but shorter than the fragmentation time. Each $\W$ can
therefore effectively be viewed as instantaneously decaying into a
string spanned between the partons. These strings expand, both
transversely and longitudinally, at a speed limited by that of light.
They eventually fragment into hadrons and disappear. Before that
time, however, the string from the $\W^+$ and the one from the
$\W^-$ may overlap. If so, there is some probability for a
colour reconnection to occur in the overlap region. The fragmentation
process is then modified.

The standard string model does not constrain the nature of the string
fully. At one extreme, the string may be viewed as an elongated bag,
i.e.\ as a flux tube without any pronounced internal structure.
At the other extreme, the string contains a very thin core, a vortex
line, which carries all the topological information, while the energy is
distributed over a larger surrounding region. The latter alternative
is the chromoelectric analogue to the magnetic flux lines in a type II
superconductor, whereas the former one is more akin to the structure
of a type I superconductor. We use them as starting points for
two contrasting approaches, with nomenclature inspired by the
superconductor analogy. In scenario~I, the reconnection
probability is proportional to the space--time volume over which the
$\W^+$ and $\W^-$ strings overlap, with strings assumed to have
transverse dimensions of hadronic size. In scenario~II, reconnections
take place when the cores of two strings cross. These two alternatives
are presented in Sections 4.4 and 4.5 respectively. As a warm-up
exercise, Section 4.3 contains a discussion of a simplified variant
of scenario~I, here called scenario~0, where strings are replaced by
simple spherical volumes.

\subsection{Relevant features of string fragmentation}

The string is the simplest Lorentz-invariant description of a
linear confinement potential. The mathematical one-dimensional
string can be thought of as parametrizing the position of the axis
of a cylindrically symmetric flux tube or vortex line. The transverse
extent of a physical string around this axis is unspecified. A string
tension of $\kappa \approx 1$ GeV/fm combined with a bag constant of
(0.23 GeV)$^4$ implies a radius of roughly 0.7 fm, i.e.\ comparable
to the proton radius.

In the decay of a $\W$, $\W^{\pm} \to \q \qbar$, a string is
stretched from the $\q$ end to the $\qbar$ one. If a number of
gluons are emitted during the perturbative phase, the string is
stretched via these gluons, i.e. from the $\q$ to the first gluon,
from there to the second one, \ldots, and from the last gluon to the
$\qbar$ end, Fig. 9 \cite{Lundstr}. The string can therefore be
described in a dual
way, either as a sequence of partons connected by string pieces,
or as a sequence of string pieces joined at gluon corners.
The gluons play the r\^ole of energy and momentum carrying kinks
on the string. Since a gluon is attached to two string pieces, the
force acting on it is twice that acting on a quark, which
always sits at the end of a string. This ratio may
be compared with the standard QCD ratio of colour Casimir factors,
$N_C / C_F = 2/(1-1/N_C^2) = 9/4$. In this, as in other
respects, the string model can be viewed as a variant of QCD
where the number of colours $N_C$ is not 3 but infinite
\cite{K8,K2}. Note that the factor of 2 above does not depend on
the kinematical configuration: a smaller opening angle between
two partons corresponds to a smaller
string length drawn out per unit time, but also to an increased
transverse velocity of the string piece, which gives an exactly
compensating boost factor in the energy density per unit string
length.

The ordering of the gluons along the string is ambiguous,
but in practice the parton shower picture, used to generate
the parton configurations, does keep track of the
colour flow and should provide a reasonable first approximation.
The string is preferentially stretched
so as to minimize the total length, i.e. partons that are nearby in
momentum space are also likely to be closely related in colour flow.
In addition to the dominant branchings $\q \to \q + \g$ and
$\g \to \g + \g$, the shower formalism also allows branchings
$\g \to \q + \qbar$. These latter split the string into two.
They will not be much covered in this paper, but
are included in the results we present.

Let us now turn to the fragmentation process, and start by considering
a $\q \qbar$ event without any energetic gluons. As the $\q$ and
$\qbar$ move apart, the potential energy stored in the string
increases, and the string may break by the production of a new
$\q' \qbar'$ pair, so that the system splits into two
colour-singlet systems $\q \qbar'$ and $\q' \qbar$. If the
invariant mass of either of these string pieces is large enough,
further breaks may occur. The string break-up process proceeds
until only on-the-mass-shell hadrons remain, each hadron corresponding
to a small piece of string with a quark at one end and an antiquark
at the other.

If transverse momenta are neglected, each break-up vertex is
characterized by two coordinates, e.g.\ $(t,z)$ for a string aligned
along the $z$ axis. Adjacent string breaks are related
by the requirement that the intermediate string piece should have the
right mass to form a hadron. Each break-up therefore effectively
corresponds to one degree of freedom. Break-ups are acausally separated,
i.e. $(\Delta t)^2 - (\Delta z)^2 < 0$, which means that there is no
unique ordering of them. They may thus be considered in any convenient
order, e.g.\ from the quark end inwards. It is therefore useful
to formulate an iterative scheme for the fragmentation, wherein
hadrons are produced one after the other in sequence, starting at
the $\q$ end. In each step the hadron carries away a fraction of
the available light-cone momentum ($E+p_z$ for a quark travelling in
the $+z$ direction), so that the remaining momentum of the string is
gradually reduced. If $m_{\perp}$ denotes the transverse mass of the
produced hadron and $\tilde{z}$ the fraction of remaining
light-cone momentum taken by the hadron, then the
$\tilde{z}$ probability distribution is given by the
`Lund symmetric fragmentation function',
\begin{equation}
f(\tilde{z}) \propto \frac{1}{\tilde{z}} \; (1-\tilde{z})^a \;
\exp(-bm_{\perp}^2/\tilde{z}) ~.
\label{Lsff}
\end{equation}
The two parameters $a$ and $b$ are to be determined from experiment.
When complemented by additional aspects, such as the generation of
transverse momenta, the appearance of different flavours and hadron
multiplets, and so on, a complete picture of the fragmentation
process is obtained \cite{Lund}.

In this classical $(1+1)$-dimensional picture, the hadron formed by
two string breaks at $(t_1, z_1)$ and $(t_2, z_2)$ (with $z_1 > z_2$
by convention) has $E = \kappa (z_1 -z_2)$ and
$p_z = \kappa (t_1 - t_2)$. Starting from the endpoint of the string,
the momenta of hadrons may then be used to recursively define
the space--time points of string breaks. The proper time $\tau$ of
string breaks therefore has an inclusive distribution, which reflects
the shape of $f(\tilde{z})$:
\begin{equation}
{\cal P}(\Gamma_{\mrm{frag}}) \propto (\Gamma_{\mrm{frag}})^a \;
\exp(- b \Gamma_{\mrm{frag}}) ~,~~~~~~\mrm{where}~
\Gamma_{\mrm{frag}} = (\kappa\tau)^2 ~.
\label{Gammafrag}
\end{equation}
Subsequent formulae become especially simple for $a \equiv 0$.
Since the best experimental values are not far away from that, we will
henceforth use $a=0$, $b \approx 0.4$ GeV$^{-2}$. (This ansatz is only
needed for the inclusive proper time distribution; it is still
possible to use a different set of $a$ and $b$ values in eq.
(\ref{Lsff}), to obtain the momenta of hadrons.) This set gives
a $\langle \Gamma_{\mrm{frag}} \rangle = (1+a)/b$ in agreement with data,
but somewhat larger fluctuations around this average than the best
experimental estimate. Introducing
$\tau_{\mrm{frag}} = 1/ \kappa b^{1/2} \approx 1.5 \mu \approx 1.5$~fm
(with $c=1$) (cf. eq. (\ref{tprodmu})), eq. (\ref{Gammafrag}) then
becomes
\begin{equation}
{\cal P}_{\mrm{frag}}(\tau) \; \d\tau =
\exp( - \tau^2/\tau_{\mrm{frag}}^2 ) \;
2 \tau  \d \tau / \tau_{\mrm{frag}}^2 ~,
\label{taufrag}
\end{equation}
where ${\cal P}_{\mrm{frag}}(\tau)$ is the differential probability that
the string will fragment at a time $\tau$.
The space--time area swept out by the string grows like $\tau^2$;
therefore an exponential decay in $\tau^2$ (rather than in $\tau$)
is to be expected when the probability for the string to break is
a constant per unit of time and length \cite{Artru}.

When the string contains several gluon kinks, the fragmentation process
is much more complicated, but can still be described in a similar
language \cite{Sjo84}. The string is breaking along its full length,
according to the same probability distribution in $\tau$ as above.
Each string piece by itself fragments into hadrons, much like a simple
$\q\qbar$ string, except at around the gluon kinks. There a hadron will
straddle the kink, i.e. contain parts of two adjacent string pieces.
Again an iterative scheme can be formulated to describe the
fragmentation from the quark ends inwards. The simple picture becomes
considerably more complicated when the invariant mass between two
adjacent partons becomes small, i.e. for soft or collinear emission.
For instance, a gluon loses its energy to the two string pieces it
pulls out in a time $t_E = E_{\g}/2\kappa$; therefore a soft
gluon with energy below roughly
$E_{\g} \approx 2\kappa\tau_{\mrm{frag}} \approx 3$ GeV
will lose its energy on a time scale shorter than the fragmentation one.
After the time $t_E$ the string motion is considerably more
complicated. We have a scheme for the momentum--energy fragmentation
process also in this case \cite{Sjo84}, but have not tried to include
the same subtleties in the space--time picture.

Clearly, the emphasis in the traditional description of string
fragmentation is on the momentum--energy picture, as in eq.
(\ref{Lsff}). A space--time
equivalent may be derived, as a by-product, but is not to be
trusted more than allowed by the uncertainty relations.
For the current paper it would have been an advantage to turn this
around, and start out from a `micro description' of the space--time
evolution. Specifically, one would have liked to trace the motion of the
string pieces and the breaking of strings by $\q\qbar$ pair creation
in time order, thereafter to translate this space--time
picture into an momentum--energy one. Our continued discussion could
then have been made more precise, in that the state of the system would
be fully specified at any potential space--time point of colour
reconnection.

The main problem with allowing string breaks in strict time order is
that it is then more difficult to simultaneously fulfil the
requirements of an overall Lorentz-invariant description and of
correct masses for hadrons. One way out is to relax the mass
constraint, by having the string break into variable-mass clusters
rather than fixed-mass hadrons. This is the approach taken in the
CALTECH-II model \cite{CTII}. However, it was never possible to
achieve a good agreement with data for CALTECH-II. In addition, there
is no clear space--time picture for the subsequent decay of clusters
into hadrons, as would be required in a complete description.

In this paper, we therefore rely on a slightly more primitive approach,
which we still think will be enough to give a good first
approximation to the more complicated full picture. The key
simplification is to divide the full process into two steps,
which are addressed in sequence rather than in parallel.
In the first step, potential colour reconnections are considered.
Here the inclusive decay distribution of eq. (\ref{taufrag}) is used
to give the probability that the string did not yet fragment by the
time of a potential recoupling, i.e.
\begin{equation}
{\cal P}_{\mrm{no-frag}}(\tau) = \int_{\tau}^{\infty}
{\cal P}_{\mrm{frag}}(\tau')
\, \d\tau' = \exp ( - \tau^2 / \tau_{\mrm{frag}}^2 ) ~.
\label{taunofrag}
\end{equation}
No other aspects of the fragmentation
process are used here. Only in the second step, after the (possibly
reconnected) string topology has been fixed, is the full machinery
of the momentum--energy picture used to fragment the strings into
an exclusive set of hadrons.

\subsection{The general space--time picture}

Consider the production of a $\W^+\W^-$ pair in the rest frame of
the process. In general, the two masses $m^{\pm} = m(\W^{\pm})$
are unequal and differ from the nominal mass $\mW$.
By momentum conservation, the absolute values of the three-momenta
agree, $p^* = |\boldp^+| = |\boldp^-|$, but the energies $E^{\pm}$
are different, $E^{\pm} = (s \pm ((m^+)^2 - (m^-)^2))/2\sqrt{s}$,
where $s = E_{\mrm{cm}}^2$ is the squared c.m. energy.
Also the boost factors, $\boldbeta^{\pm} = \boldp^{\pm}/E^{\pm}$
and $\gamma^{\pm} = E^{\pm}/m^{\pm}$, therefore differ.

In a complete description of the production process, there is a
competition between the phase space and the $\W^{\pm}$
Breit--Wigners. (Also the form of the matrix element (including
Coulomb final state interactions) plays a r\^ole, although we may
neglect that for qualitative considerations.) For
$E_{\mrm{cm}} = 2 \mW$ this results in a
$\langle p^* \rangle \approx 22$ GeV, rather than
the na\"{\i}ve $p^* = 0$. The $\langle p^* \rangle$ does change
with c.m. energy, but less rapidly than in the na\"{\i}ve picture,
Fig. 10a. Above $2 \mW$ the competition gradually
becomes less important; for $E_{\mrm{cm}} = 200$~GeV the
$\langle p^* \rangle$ is just what one would expect
for $\W$'s on the mass shell. If instead the c.m. energy is decreased
below $2 \mW$, at least one $\W$ can no longer benefit from the
Breit--Wigner peak enhancement. Since the Breit--Wigner varies less
rapidly in the tails, the phase space factor becomes more important,
proportionally speaking. Therefore $\langle p^* \rangle$ is also
increased at lower $E_{\mrm{cm}}$.

The average proper lifetime of a $\W$ depends on its mass $m$
according to
\begin{equation}
\langle \tau \rangle = \tau_{\mrm{dec}}(m)
= \frac{\hbar m}{\sqrt{(m^2 - \mW^2)^2 +
(\GW m^2/\mW)^2}} ~,
\label{Wlifetime}
\end{equation}
where $1 = \hbar \approx 0.197$ GeV$\cdot$fm.
For a $\W$ on the mass shell this reduces to the standard experession
$\tau_{\mrm{dec}}(\mW) = \hbar/\GW \approx 0.1$ fm. However,
with a standard
Breit--Wigner distribution of masses, typically the $\W$ lifetime is
only about two thirds as long as the na\"{\i}ve expectation, Fig. 10b.
The $\W$ width $\GW \approx 2.1$ GeV has been defined for
$m = \mW$; the variation of the width as a function of mass has
been included as an explicit factor $m/\mW$ in eq.
(\ref{Wlifetime}).

The actual proper lifetime of a $\W^{\pm}$ is thus distributed
according to
\begin{equation}
{\cal P}(\tau^{\pm}) \, \d\tau^{\pm} = \exp ( - \tau^{\pm} /
\tau_{\mrm{dec}}(m^{\pm}) ) \; \d\tau^{\pm} /
\tau_{\mrm{dec}}(m^{\pm}) ~.
\label{Wlifedist}
\end{equation}
If the $\W^+\W^-$ pair is created at the origin,
$(\boldx_0^0,t_0^0) = (\boldzero,0)$, the $\W^{\pm}$ decay vertices
are given by $(\boldx_0^{\pm}, t_0^{\pm}) =
(\gamma^{\pm}\boldbeta^{\pm}\tau^{\pm}, \gamma^{\pm}\tau^{\pm})$.
The average separations $|t_0^+ - t_0^-|$ and
$|\boldx_0^+ - \boldx_0^-|$ as a function of the c.m. energy are
shown in Fig. 10b. At typical LEP 2 energies the time separation is
about 0.08 fm and the spatial separation 0.05 fm.

Each $\W$ decays to a $\q\qbar$ pair. The quarks normally are off the
mass shell and therefore branch further, $\q \to \q + \g$. The
daughter partons may branch in their turn, and so on. A parton shower
thus develops, to leading order made up out of the three branchings
$\q \to \q + \g$, $\g \to \g + \g$ and $\g \to \q + \qbar$. The
branchings are ordered in mass, by trivial kinematical constraints,
i.e. daughter partons have to be less virtual than the
mother parton. In current parton shower algorithms, the evolution is
stopped at some lower cut-off scale, typically $m_0 \approx 1$ GeV.
Branchings may well occur at lower scales, but can no longer be
described in perturbative terms; they are instead effectively
included in the fragmentation description. The branching process is
ambiguous, in the sense that one given partonic final state may
be arrived at by a host of intermediate branching histories.
However, coherence effects impose a further ordering in terms of
decreasing emission angles \cite{K2}, which limits this
ambiguity. Several shower algorithms have been proposed, which differ
in technical details, but agree in most of their predictions. (An
exception is prompt photon production, which may offer an opportunity
to learn more about shower evolution \cite{promptgamma}.) In the
following, we use the one of ref. \cite{MBTS}.

It is not unreasonable to neglect on-shell quark masses, since the
heaviest quark produced with any significant rate is the $\c$ one
($\W^- \to \b \cbar$ decays are negligible). The average lifetime
of a parton in a shower is then given by its off-shell mass $m$ and
energy $E$:
\begin{equation}
\langle t_{\mrm{part}} \rangle =
\gamma \, \langle \tau_{\mrm{part}} \rangle =
\frac{E}{m} \, \frac{\hbar}{m} = \frac{\hbar E}{m^2} ~.
\end{equation}
A parton with a mass close to the lower cut-off,
$m \approx m_0 \approx 1$ GeV, and a maximal energy,
$E \approx \mW/2 \approx 40$ GeV, would thus have time to travel
about 8 fm before branching to the final partons. This is a distance
much larger than the separation between the $\W^+$ and $\W^-$ decays,
and so it is to be feared that it is important to keep track of the
space--time evolution of the shower. However, branchings at such low
mass scales do not give rise to separate jets, but only to some
additional transverse momentum smearing inside the hadron fragments
of a jet. At LEP 1, the limit for meaningful jet resolution is
typically $m \approx 10$ GeV
($y_{ij} = m_{ij}^2/E_{\mrm{cm}}^2 > y_{cut} \approx 0.01$), which
corresponds to $\langle t_{\mrm{part}} \rangle \approx 0.1$ fm.
The overall event structure is thus determined on a scale comparable
to the separation between the $\W^+$ and $\W^-$ decays, and at a time
much shorter than typical fragmentation scales. For the subsequent
discussion of string motion we will thus assume that all partons of a
$\W$ decay have a common origin. The effects of the
low-mass branchings will be studied by cutting off the shower at
different $m_0$ scales.

As the partons move apart they pull out a string, made up of
straight string pieces between adjacent partons.
The most general case we need to consider is a string piece created at
a point $(\boldx_0,t_0)$, with the two endpoints of the string moving
out with velocities $\boldv_1 = \boldp_1/E_1$ and
$\boldv_2 = \boldp_2/E_2$. Usually we will neglect the possibility
that the endpoint partons are off the mass shell, i.e. assume that
$|\boldv_1| = |\boldv_2| = 1$. The string position can then
be described as
\begin{equation}
\boldx_{\mrm{string}}(t) = \boldx_0 + (\boldv_1 + \alpha (\boldv_2 -
\boldv_1)) \; (t - t_0) ~,~~~~~~0 \leq \alpha \leq 1 ~,
\label{stringmove}
\end{equation}
with the centre of the string at $\alpha = 1/2$.
Therefore the overall motion of the string is given by
$\boldbeta = (\boldv_1 + \boldv_2)/2$ and a unit vector along the
string direction by
$\boldu = (\boldv_2 - \boldv_1)/|\boldv_2 - \boldv_1|$.
One obtains
$|\boldbeta| = \sqrt{(1+\cos\theta_{12})/2} = \cos(\theta_{12}/2)$ and
$\gamma = 1/\sin(\theta_{12}/2)$, where $\theta_{12}$ is the
angle between the two partons. The $\gamma$ factor gives the time
dilatation for the fragmentation process in the middle of the string,
i.e. $\langle t_{\mrm{frag}} \rangle = \gamma \tau_{\mrm{frag}}$.
The regions of the string closer to the endpoints of the string
obviously fragment later, on the average.

We saw above that high-virtuality partons decay in times much shorter
than typical fragmentation times. Low-mass partons, on the other hand,
can travel distances larger than $\tau_{\mrm{frag}}$, and so it is of some
interest to study the importance of boost effects on the fragmentation
process. Consider a quark or gluon with an energy $E$ and mass $m$,
where the two daughters take energy fractions $z$ and $1-z$,
respectively. The new string piece produced by the parton branching
is spanned between these two daughters. The opening angle
$\theta_{12}$ can be approximated by
\begin{equation}
\theta_{12} = \theta_1 + \theta_2 \approx
\frac{\pT}{zE} + \frac{\pT}{(1-z)E} = \frac{\pT}{z(1-z)E}
\approx \frac{m}{\sqrt{z(1-z)} E} ~,
\end{equation}
where $\theta_i$ is the angle of either daughter and $\pT$ is their
common transverse momentum with respect to the mother direction.
Therefore
\begin{equation}
\langle t_{\mrm{frag}} \rangle = \gamma \tau_{\mrm{frag}} =
\frac{\tau_{\mrm{frag}}}{\sin(\theta_{12}/2)} \approx
\frac{2\tau_{\mrm{frag}}}{\theta_{12}} \approx
2\sqrt{z(1-z)} \, \frac{E}{m} \, \tau_{\mrm{frag}} ~,
\end{equation}
and
\begin{equation}
\frac{\langle t_{\mrm{frag}} \rangle}{\langle t_{\mrm{part}} \rangle}
\approx \frac{2\sqrt{z(1-z)} E \tau_{\mrm{frag}}/m}{\hbar E/m^2} =
\frac{2}{\hbar} \sqrt{z(1-z)} \, m \, \tau_{\mrm{frag}} \approx
(15~\mrm{GeV}^{-1}) \sqrt{z(1-z)} \, m ~.
\end{equation}
It is easy to convince oneself that this ratio is comfortably bigger
than unity in the whole physical region. In the extreme case of
$m = m_0 \approx 1$ GeV and
$z = E_1/E = (m_0/2)/(\mW/2) \approx 1/100$, one obtains
$\langle t_{\mrm{frag}} \rangle / \langle t_{\mrm{part}} \rangle
\approx 1.5$.

The difference between the correct string drawing and the one obtained
by setting $t_{\mrm{part}}=0$ is shown in Fig. 11 for a very simple
example. The $\W$ decays to a
$\q^*\qbar$ pair at point A. The $\q^*$ is virtual, and subsequently
branches, $\q^* \to \q + \g$, at point B. The correct string topology
at some later time is shown by Fig. 11a. Closest to the $\qbar$ end is
a string piece pulled out by the $\q^*$ and the $\qbar$ before the
$\q^*$ branched, and this piece is therefore aligned along the
original $\q^*\qbar$ event axis. Next comes the string piece pulled
out by the $\qbar$ and the $\g$ after the latter was produced. The
kink (C) that joins these two string pieces does not carry any energy,
unlike an ordinary gluon kink. It was formed at the $\q^*$ decay vertex
and is travelling in the $\qbar$ direction with the speed of light.
Finally, there is the string piece between the $\g$ and the $\q$,
expanding from point B. Figure 11b shows the same picture in the
approximation that point B coincides with A, so that there is no
string piece pulled out between $\q^*$ and $\qbar$.

\subsection{Scenario 0: spherical volumes}

Strings are assumed to have a well-defined longitudinal direction.
However, if the main jets of the $\W^+$ and the $\W^-$ are well
separated, the strings only overlap in the middle.
To first approximation, it is therefore useful to consider
the case of a string being a completely spherical colour source.
The probability for a reconnection to occur is taken to be proportional
to the overlap of the $\W^+$ colour source with the $\W^-$ one, with
details to be specified in the following.

The result is a very simple model for colour reconnection,
which gives some feeling for the energy dependence of the reconnection
probability, and where several results can be obtained without recourse
to a complete event generator. This toy model is thus not of
comparable scope with the two other scenarios we will develop below.
Clearly, any information on the angular
distributions of the $\W^{\pm}$ decays is lost.
In particular, if one jet from the $\W^+$ and one from the $\W^-$ move
out in the same general direction the strings will overlap for longer,
and the spherical approximation is likely to break down.

The colour field strength $\Omega_0$ of a spherically symmetric colour
source at rest at the origin can be approximated by
\begin{equation}
\Omega_0(\boldx,t)
= \exp (- \boldx^2/2r_{\mrm{had}}^2) \; \theta(t - |\boldx|) \;
\exp (- t^2/\tau_{\mrm{frag}}^2) ~.
\label{Omegazero}
\end{equation}
The first factor corresponds to a Gaussian fall-off of the field
strength. Alternatives could certainly be considered, with a more or
less sharp edge, but would not affect the qualitative picture.
We have seen above that the typical transverse size of a string is
$r_{\mrm{cyl}} \approx 0.7$ fm when the string is described as a
cylinder with a sharp edge. If this is replaced by a Gaussian
fall-off in two transverse dimensions, an appropriate choice of
width in each dimension
is $r_{\mrm{had}} \approx r_{\mrm{cyl}}/\sqrt{2} \approx 0.5$ fm.
The subsequent
step function $\theta(t - |\boldx|)$ ensures that information on the
decay of the $\W$, assumed to take place at $t=0$, spreads outwards
at the speed of light ($c=1$). The final factor is the probability
that a string remains at time $t$, i.e. has not yet fragmented, eq.
(\ref{taunofrag}). The typical proper lifetime is
$\tau_{\mrm{frag}} \approx 1.5~\mrm{fm} \approx 3 r_{\mrm{had}}$.
Some time-retardation could be included also here, but then it would
be necessary to first specify the spatial point of fragmentation,
which would mean taking the model more seriously than it warrants.

In a second step, we
generalize to a moving colour source, again created at the origin
but moving away with a speed $\boldbeta$. The evaluation of
$\Omega(\boldx,t)$ is most conveniently done by performing a
boost $-\boldbeta$ back to the rest frame of the source:
\begin{eqnarray}
\Omega(\boldx,t;\boldbeta) & = & \Omega_0(\boldx',t')~, \\
\boldx' & = & \boldx + \gamma \left(
\frac{\gamma\boldbeta\boldx}{1+\gamma} - t \right) \boldbeta ~,
\label{xprime} \\
t' & = & \gamma (t - \boldbeta\boldx)
\label{tprime}
\end{eqnarray}
(remember that the volume element
$\d^3\boldx \, \d t = \d^3\boldx' \, \d t'$ is boost-invariant).
Finally, if the source is not created at the
origin but at a point $(\boldx_0, t_0)$, the distribution is
\begin{equation}
\Omega(\boldx,t; \boldx_0,t_0; \boldbeta) =
\Omega(\boldx - \boldx_0, t - t_0; \boldbeta) =
\Omega_0((\boldx - \boldx_0)', (t - t_0)')~.
\end{equation}

Now consider the production of a $\W^+\W^-$ pair at the origin,
moving out back-to-back. In this simple toy study we
assume both $\W$'s to have the same nominal mass $m = \mW=80$ GeV.
The common velocity is therefore
$\beta = |\boldbeta| = \sqrt{1 - 4\mW^2/s}$.
The proper lifetime of each $\W$ is distributed according to
eq. (\ref{Wlifedist}) with $\tau_{\mrm{dec}} = \tau_{\mrm{dec}}(\mW)$.
The overlap of the two sources, averaged over the $\W^{\pm}$ lifetime
spectra, is then given by
\begin{equation}
{\cal I}(\beta) = \int {\cal P}(\tau^+) \; \d\tau^+
\int {\cal P}(\tau^-) \; \d\tau^- \int \d^3\boldx \, \d t \;
\Omega(\boldx,t;\gamma\tau^+\boldbeta,\gamma\tau^+) \;
\Omega(\boldx,t;-\gamma\tau^-\boldbeta,\gamma\tau^-)~.
\label{OOmega}
\end{equation}

Since the absolute normalization of ${\cal I}(\beta)$ is irrelevant,
results are conveniently normalized to ${\cal I}(0)$, which is
the maximum value. This ratio ${\cal I}(\beta)/{\cal I}(0)$ is
shown in Fig. 12. There are a few comments to be
made. Firstly, the variation of ${\cal I}(\beta)$ with c.m. energy
is rather slow. In other words, the `threshold region' of potentially
large reconnection probabilities covers the whole LEP 2 range, with
a variation in ${\cal I}(\beta)$ of a factor 3--4.
Secondly, the curves show that this variation is
shared between three contributing factors: the motion of the sources,
the time retardation factor, and the decay of $\W^{\pm}$ away from
the origin. Of these, the last one is the least important, in spite
of us having used the long lifetime of on-the-mass-shell $\W$'s.
With a realistic $\W$ mass spectrum, the displacement of the $\W$ decay
vertices is therefore even less significant.

We may assume the probability for string recoupling to be
proportional to ${\cal I}(\beta)$, with some unknown constant
of proportionality. However, should the probability become large,
saturation must set in. In the current scenario only two configurations
are possible, $\q_1\qbar_2 + \q_3\qbar_4$ and
$\q_1\qbar_4 + \q_3\qbar_2$, so each recoupling corresponds to
flipping between these two. The probability for the latter
configuration should therefore exponentially approach a saturation
value of $1/2$ , i.e.
\begin{equation}
{\cal P}_{\mrm{recon}} = {\cal P}(\q_1\qbar_4 + \q_3\qbar_2) =
\frac{1}{2} \; \left( 1 - \exp \left( -k_0 \frac{{\cal I}(\beta)}%
{{\cal I}(0)} \right) \right) ~.
\label{reconsphere}
\end{equation}
Depending on the $k_0$ value chosen one may obtain widely different
results for ${\cal P}_{\mrm{recon}}$, see Fig. 13. Obviously, the
energy variation of  ${\cal P}_{\mrm{recon}}$ is never faster than
that of ${\cal I}(\beta)$.

To the list of uncertainties already mentioned, one should add the
possibility of a modified velocity dependence in eq.
(\ref{OOmega}). As it stands, this equation only gauges the
geometrical overlap of two colour sources, but does not specify the
mechanism whereby the colour reconnection occurs. Rapidly moving
colour sources might interact differently than ones at rest,
presumably more intensely, such that the net variation of
reconnection probability with c.m. energy could be even slower
than shown in Fig. 13.

In summary, the main lesson of this simple exercise is that the colour
reconnection phenomenon is likely to have a very extended threshold
region. An increase of LEP~2 energy could not be used to make an
`undesirable' phenomenon `go away'. Let us recall that, also in the
perturbative scenario, the magnitude of the reconnection effects
(albeit small) is expected to be comparable at the threshold and
reasonably far above it ($E_{\W} \sim {\cal O}(\mW)$).

\subsection{Scenario I: elongated bags}

In this scenario strings are assumed to be (time-retarded) cylindrical
bags, and the recoupling probability to be proportional to the
integrated overlap between such cylinders. The formalism  has many
similarities with the preceding one, but is intended to be more
realistic, in a number of respects.

If a string, viewed in its rest frame, is expanding along the
direction $\pm\boldu$, $|\boldu|$ = 1, the colour field strength
$\Omega_0$ may be written as
\begin{equation}
\Omega_0(\boldx,t;\boldu) =
\exp \left\{ - (\boldx^2 - (\boldu\boldx)^2)/2r_{\mrm{had}}^2 \right\}
\; \theta(t - |\boldx|) \;
\exp \left\{ - (t^2 - (\boldu\boldx)^2)/\tau_{\mrm{frag}}^2 \right\} ~.
\label{OmegazeroI}
\end{equation}
As for eq. (\ref{Omegazero}), the first term gives a Gaussian
fall-off, but now only in the transverse directions (if
$\boldu = (0,0,1)$ then
$\boldx^2 - (\boldu\boldx)^2 = r^2 - z^2 = x^2 + y^2$). The
time retardation factor $\theta(t - |\boldx|)$ is unchanged.
The last factor, the probability that the string has not yet
fragmented, now depends on the proper time along the string axis
($\tau^2 = t^2 - (\boldu\boldx)^2 = t^2 - z^2$ for a string along the
$z$ axis).

Now consider a moving string piece, created at
a point $(\boldx_0,t_0)$, with the string position described as in
eq. (\ref{stringmove}). The transverse motion direction of the
string piece is given by $\boldbeta$ and a unit vector along the
string direction by $\boldu$. The colour field strength in an
arbitrary point $(\boldx,t)$ is then
\begin{equation}
\Omega(\boldx,t;\boldx_0,t_0;\boldbeta;\boldu) =
\Omega(\boldx - \boldx_0, t - t_0;\boldbeta;\boldu) =
\Omega_0((\boldx - \boldx_0)', (t - t_0)'; \boldu) ~,
\end{equation}
where boosted coordinates $\boldx'$ and $t'$ are given by eqs.
(\ref{xprime}) and (\ref{tprime}). Note that $\boldu$ is unchanged
by the boost, $\boldu' = \boldu$, since $\boldbeta$ and $\boldu$
are orthogonal and $\boldu$ is a pure space-like vector
($v_i = (\boldv_i, 1) \Rightarrow u = (\boldu,0)$).

Had perturbative QCD effects been neglected, each $\W$ would have
decayed to a $\q\qbar$ pair, i.e. one single string piece, with the
two endpoint vectors $\boldv_i$ given by the $\q$ and $\qbar$.
In the full description, however, the perturbative parton shower
produces a number of gluons, and the string is therefore stretched
from the $\q$ via these gluons to the $\qbar$. A string with $n$
partons, labelled
$\q_1 \, \g_2 \, \g_3 \, \cdots \, \g_{n-1} \, \qbar_n$,
has $n-1$ separate string pieces. Each of these gives a colour source
$\Omega(\boldx,t;\boldx_0,t_0;\boldbeta_i;\boldu_i)$,
where $(\boldx_0,t_0)$ is the common decay vertex of the $\W$, while
the $\boldbeta_i$ and $\boldu_i$ are defined by the
direction vectors $\boldv_i$ and $\boldv_{i+1}$ of the two adjacent
partons along the string.

Normally at most one of the many pieces of a string should contribute
to the colour field in any specific space--time point. However, with
the simple ansatz of eq. (\ref{OmegazeroI}), there is no cut-off to
stop the overlapping of several colour fields. Such overlaps are
especially likely to happen close to a gluon corner, where two pieces
meet, but also occur because our simple transverse Gaussians in
principle extend all the way to infinity. Where overlaps do occur,
normally one field is much more important than the rest. As a simple
solution, the colour field strength of the decay products of a $\W$
is therefore defined as the maximum of all possible contributions,
\begin{equation}
\Omega_{\mmax}(\boldx,t;\boldx_0,t_0) = \max_{i=1,n-1}
\Omega(\boldx,t;\boldx_0,t_0;\boldbeta_i;\boldu_i) ~.
\label{Omegamaxdef}
\end{equation}

In a $\W^+\W^-$ event, the overlap between the colour fields of
the $\W^+$ decay products, on the one hand, and the $\W^-$ ones, on
the other, is thus given by
\begin{eqnarray}
{\cal I}^{+-}
& = & \int \d^3\boldx \, \d t \;
\Omega_{\mmax}^+(\boldx,t;\boldx_0^+,t_0^+) \;
\Omega_{\mmax}^-(\boldx,t;\boldx_0^-,t_0^-) \nonumber \\
& = & \int \d^3\boldx \, \d t \;
\Omega_{\mmax}^+(\boldx,t;\gamma^+\tau^+\boldbeta^+,\gamma^+\tau^+) \;
\Omega_{\mmax}^-(\boldx,t;\gamma^-\tau^-\boldbeta^-,\gamma^-\tau^-) ~.
\label{OOmegaI}
\end{eqnarray}

The integral ${\cal I}^{+-}$ is too complicated to allow any
analytical short-cuts, and has to be evaluated numerically for each
individual event, since it explicitly depends on the specific
parton configuration. This would be very time-consuming if good
accuracy is strived for. Fortunately it is possible to take a
numerical shortcut. If a Monte Carlo method is used to estimate
the integral, then, to first approximation, the estimated
${\cal I}^{+-}$ value need not be the right one for each separate
event, as long as the average is correct for an imagined infinite
ensemble of identical events. It could be argued that, to second
approximation, an error is made, since the reconnection probability
${\cal P}_{\mrm{recon}}$ should saturate and therefore not be quite
proportional to ${\cal I}^{+-}$, cf. eq. (\ref{reconsphere}).
However, in scenario 0 there was only one possible colour
reconnection, while there is here a multitude of them, depending on
which string pieces are involved. Saturation should therefore be
related more to the local overlap of $\Omega_{\mmax}^+$ and
$\Omega_{\mmax}^-$ than to the overlap integral over the
full space--time. Further, the simple classical colour field
configurations that we have introduced should be seen as smoothed-out
averages of a more `grainy' structure. Therefore we can make virtue
of necessity, and use the Monte Carlo approach as a way to introduce
a more realistic interpretation of the colour field.

In this spirit, the integral is approximated by
\begin{eqnarray}
{\cal I}^{+-} & = & \int \d^3\boldx \, \d t \;
\Omega_{\mmax}^+(\boldx,t;\boldx_0^+,t_0^+) \,
\Omega_{\mmax}^-(\boldx,t;\boldx_0^-,t_0^-)
\nonumber \\
& = & \int \Omega_{\mrm{trial}}(\boldx,t) \, \d^3\boldx \, \d t \;
\frac{\Omega_{\mmax}^+(\boldx,t;\boldx_0^+,t_0^+) \,
\Omega_{\mmax}^-(\boldx,t;\boldx_0^-,t_0^-)}
{\Omega_{\mrm{trial}}(\boldx,t)}
\nonumber \\
& \propto & \left\langle \frac{\Omega_{\mmax}^+\Omega_{\mmax}^-}
{\Omega_{\mrm{trial}}}(\boldx,t;\boldx_0^+,t_0^+;\boldx_0^-,t_0^-)
\right\rangle
\nonumber \\
& \approx & \frac{1}{100} \, \sum_{i=1}^{100}
\frac{\Omega_{\mmax}^+\Omega_{\mmax}^-}{\Omega_{\mrm{trial}}}
(\boldx_i,t_i;\boldx_0^+,t_0^+;\boldx_0^-,t_0^-)
\equiv {\cal I}_{\Sigma}^{+-} ~.
\label{Omegawappr}
\end{eqnarray}
In the last two lines it is to be understood that the points
$(\boldx_i,t_i)$ are to be selected at random according to the trial
distribution $\Omega_{\mrm{trial}}$, which has been taken as
\begin{equation}
\Omega_{\mrm{trial}}(\boldx,t) =
\exp(- \boldx^2/ f_r^2 r_{\mrm{had}}^2) \;
\exp(- 2 (t - t^{+-})^2/f_t^2 \tau_{\mrm{frag}}^2) \;
\theta (t - t^{+-}) ~,
\end{equation}
with $t^{+-} = \max(\gamma^+\tau^+, \gamma^-\tau^-)$.
The limit $f_r, f_t \to \infty$ corresponds to a `correct' sampling
of the full space--time, with vanishing probability to sample an
interesting point. The values $f_r = 2.5$ and $f_t =2$ give a
compromise, which allows reasonable efficiency without introducing
too much bias of saturation effects (see below). Results are not
sensitive to this choice.

Out of the 100 space--time points selected for each event, only about
30 give a non-vanishing contribution; the others are outside the
forward light-cone either of the $\W^+$ string or of the $\W^-$ one.
The sum of weights in eq. (\ref{Omegawappr}) is used to determine
whether a recoupling should occur or not. In case of recoupling, one
of the possible points is selected, with a probability proportional
to the weight of this point. The selected point is associated with
the overlap of a specific string piece from the $\W^+$ with another
from the $\W^-$, any potential ambiguities having been resolved by
the recipe of eq. (\ref{Omegamaxdef}). Thus it is possible also to
select where on the two strings the reconnection occurs.

With reconnection possible in several different string pieces,
it is not obvious at what value the reconnection probability should
be forced to saturate, since two reconnections do not have to
cancel each other. To see this, consider two strings
$\q^+\g^+\qbar^+$ and $\q^-\g^-\qbar^-$ from the $\W^{\pm}$ decays.
A first reconnection between the $\q^+\g^+$ and $\g^-\qbar^-$ string
pieces would give the two new strings $\q^+\qbar^-$ and
$\q^-\g^-\g^+\qbar^+$. A second reconnection between the
$\q^-\g^-$ and $\g^+\qbar^+$ pieces would not give back the
original configuration, but rather lead to a state consisting of
three strings, $\q^+\qbar^-$, $\q^-\qbar^+$ and $\g^+\g^-$.
(Multiple string reconnections could be an interesting source of pure
gluonium states. Also a single string, e.g.\ in $\Z^0$ decay, could
fold back on itself in such a way that a gluon loop could be split
off. This could be a production mechanism, e.g.\ for
$\JP$ and $\Upsilon$ states.)

For simplicity, we have chosen to allow at most one reconnection
per event, but to set saturation at unit reconnection probability,
i.e.\
\begin{equation}
{\cal P}_{\mrm{recon}} =  1 - \exp \left( - k_{\mrm{I}} \, f_r^3 f_t \,
{\cal I}_{\Sigma}^{+-} \right) ~.
\label{reconcylin}
\end{equation}
Note that ${\cal I}_{\Sigma}^{+-}$ is somewhere in between a local
and a global quantity, and therefore represents a pragmatic
compromise between the conflicting demands. When all weights
in the sum are of moderate size, saturation is no problem, and
the na\"{\i}ve behaviour is recovered, i.e.\ the reconnection
probability in a point is given by the
overlap $\Omega_{\mmax}^+\Omega_{\mmax}^-$
in that point. If one weight is very large,
the averaging procedure with the other weights somewhat smoothens
fluctuations in ${\cal P}_{\mrm{recon}}$. Once the decision has been
taken to allow a reconnection, however, the large weight does
have its full impact on the choice of where this reconnection
occurs. The above procedure contains one obvious error, namely that
the saturation introduces an artificial dependence on the choice
of $\Omega_{\mrm{trial}}$. This tends to reduce the importance of
recouplings at large distances and times, but not enough to provide
a serious distortion of results.

A priori, the constant $k_{\mrm{I}}$ in eq. (\ref{reconcylin}) is
unknown. Unlike the spherical case, there is no ${\cal I}(0)$
to normalize to, so we only explicitly separate out the dependence
on the choice of $\Omega_{\mrm{trial}}$. Instead $k_{\mrm{I}}$ will later
be determined by comparison with scenario II below.

\subsection{Scenario II: vortex lines}

In scenario II it is assumed that strings are like vortex lines
in type II superconductors. Then all the topological information
is given by the small core region, even if most of the energy is
stored in the region surrounding the core. Reconnections can only
take place when the core regions of two string pieces cross
each other. This means that the transverse extent of
strings can be neglected, which leads to considerable
simplifications compared with the previous scenario. It is every bit
as well founded physically as scenario~I.

Consider string piece $i$ spanned between partons $i$ and $i+1$
of the $\W^+$ string, and string piece $j$ spanned between
partons $j$ and $j+1$ of the $\W^-$ string. By analogy with eq.
(\ref{stringmove}), the position of the two string pieces may
be described as
\begin{eqnarray}
\boldx_i^+(t , \alpha^+) & = & \boldx_0^+ + (\boldv_i + \alpha^+
(\boldv_{i+1} - \boldv_i)) \; (t - t_0^+) ~,
{}~~~~~~0 \leq \alpha^+ \leq 1 ~,
\nonumber \\
\boldx_j^-(t , \alpha^-) & = & \boldx_0^- + (\boldv_j + \alpha^-
(\boldv_{j+1} - \boldv_j)) \; (t - t_0^-) ~,
{}~~~~~~0 \leq \alpha^- \leq 1 ~.
\end{eqnarray}
To find whether the two string pieces cross, one could step through
the string evolution as a function of time. However, it is more
convenient to solve
\begin{equation}
\boldx_i^+(t , \alpha^+) = \boldx_j^-(t , \alpha^-) ~.
\label{stringcross}
\end{equation}
This is a system of three equations --- $x$, $y$ and $z$ ---
with three unknowns --- $t$, $\alpha^+$ and $\alpha^-$.
If the values of the unknowns are not constrained, the system
always has exactly one solution, except for trivial pathologies
such as two strings completely at rest or exactly parallel to each
other. The uniqueness of the solution is easy to understand, by
simple geometrical considerations for two moving, infinitely long
strings. To find out if two string pieces actually do cross each
other, it is therefore only necessary to check whether the solution
is in the physical domain $t > \max(t_0^+,t_0^-)$,
$0 \leq \alpha^+ \leq 1$, and $0 \leq \alpha^- \leq 1$.

To the above requirements comes the additional constraint that
neither string piece must have had time to fragment before the
time of crossing. In agreement with our earlier assumption,
this is taken to be given by a Gaussian fall-off in the proper
time of each string piece at the point of crossing,
\begin{equation}
{\cal P}_{\mrm{no-frag}} = \exp \! \left( - \frac{ (t - t_0^+)^2 -
(\boldx_i^+(t , \alpha^+) - \boldx_0^+)^2}{\tau_{\mrm{frag}}^2}
\right) \; \exp \! \left( - \frac{ (t - t_0^-)^2 -
(\boldx_j^-(t , \alpha^-) - \boldx_0^-)^2}{\tau_{\mrm{frag}}^2}
\right) ~.
\end{equation}

If a string crossing occurs, according to the above criteria,
then a reconnection is assumed to take place with unit probability.

This scenario does not have any free parameters at all, except for
$\tau_{\mrm{frag}}$, of course, which may be considered as a known
constant. It is therefore much more constrained and predictive
than the ones encountered above. However, it would be wrong to
leave the impression that everything is understood. Below we give
a list of some uncertainties that one should be aware of, and also
of some potential variations from the basic model.
\begin{Itemize}
\item
Occasionally an event may allow several string crossings, which could
then, for instance, give rise to closed gluon loops, as discussed
before. In the current study, at most one reconnection has been
allowed, namely the one (among the potential ones) that occurs
first in time. We do not expect the inclusion of some additional
reconnections to have any influence on the studies presented in
this paper.
\item
The gain in string length by a reconnection depends on
the topology of the event, as illustrated in Fig. 14. (For string
pieces spanned between gluons, a $\g$ is represented by a $\q$
or a $\qbar$, depending on the direction of the colour flow along
the string.) In a configuration like the one in Fig. 14a, the change
in string topology is drastic, so a reconnection would almost
certainly occur. If the configuration is more like Fig.~14b,
where the two $\q$ (and/or $\qbar$) ends are close together, a
reconnection does not imply a major change. If quantum mechanical
fluctuations are introduced around the classical limit, one might
then expect the reconnection probability to be close to $1/2$.
It is also possible to have crossings where the string length would
be increased by a reconnection, such as Fig. 14c, and the
reconnection probability consequently could be smaller than $1/2$.
Crossings which give a large increase in string length are not
favoured geometrically, but some 20\% of all crossings correspond
to an increase rather than a decrease (at 170 GeV).
Additional parameters would be needed to model a smooth change
in the reconnection probability with topology, without necessarily
adding new insights. Therefore we have not pursued it,
except to study an option where recouplings are allowed only
if they reduce the string length. For two $\q_1\qbar_2$ and
$\q_3\qbar_4$ (or equivalents with gluons) string pieces, this
translates into the requirement that $m_{14}m_{23} < m_{12}m_{34}$,
where $m_{ij}$ is the invariant mass of partons $i$ and $j$.
We call this scenario II$'$.
\item
It could be argued that strings with different endpoint colours
could pass through each other without interacting, so that
the interaction probability would be further reduced. Based on the
discussion in section 3.6 we choose to neglect this possibility.
(The same argument could be raised
for scenarios 0 and I, although colour is more smeared-out in those
two, so that the reduction is more likely to be in the parameters
$k_0$, $k_{\mrm{I}}$ than in ${\cal P}_{\mrm{recon}}$ itself.
Thus nothing new is introduced, given that $k_0$, $k_{\mrm{I}}$
are unknown anyway.)
\item
The more gluon emission is occurring in the parton shower evolution,
the more a string will twist around in space, and the more likely
it is that the strings from the two $\W$'s intersect. Results are
therefore sensitive
to the amount of soft-gluon emission, i.e.\ to the $m_0$ cut-off in
the parton shower. This sensitivity will be studied later on.
Also scenario I has a dependence on the partonic state, but not
to the same extent: a tube around a string core is a smoother
object than the core itself.
\item
The approach of letting all string regions expand away from the
$\W$ vertex, as we do, is not correct.
Low-virtuality partons travel out a long distance before they
branch. This means that some string regions will not be formed
until after the system has spread outwards enough for string
crossings to be rare. However, such string pieces generally subtend
small angles, and are therefore not likely to be the ones that
cross in the first place. Further, before the parton branched, it
is not that a given angular region was empty of strings; that
region was still filled, only by a string with one less kink.
This uncertainty is therefore intimately related to the soft-gluon
issue above.
\end{Itemize}

\section{Non-perturbative model predictions}

In this section we present results based on a complete simulation
of the non-perturbative scenarios I and II described in Section 4.

\subsection{Reconnection probability}

The reconnection probability ${\cal P}_{\mrm{recon}}$ is predicted in
scenario II without any adjustable parameters, although with
uncertainties as already noted. Scenario I contains a completely
free strength parameter $k_{\mrm{I}}$, which could be adjusted to
give any desired ${\cal P}_{\mrm{recon}}$. To ease comparisons between
the two alternatives, we have chosen $k_{\mrm{I}} = 0.6$, such that
the rates of reconnected events agree at $E_{\mrm{cm}} \approx 170$ GeV.

The resulting energy dependence of ${\cal P}_{\mrm{recon}}$ is shown in
Fig. 15a. The main message is that the variation is very slow:
the reconnection probability varies by at most a factor of 2 over
the c.m. energy range 150--200 GeV. This is much slower than in
scenario~0, see Fig. 13. The key difference is that scenario 0
did not include any Breit--Wigners for the $\W$ masses, so that
the two $\W$'s were produced at rest for $E_{\mrm{cm}} = 2 \mW$.
In scenarios I and II, on the other hand, the effects of Breit--Wigners
are included. As shown in Fig. 10a, this leads to a less rapid
variation of $\langle p^* \rangle$. A larger $\langle p^* \rangle$
gives $\W^+$ and $\W^-$ decay points that are more separated, Fig. 10b,
and $\W$ decay products that are faster moving away from each other.
The variation of the reconnection probability is thus directly related
to that of the $\langle p^* \rangle$. Below $2\mW$ the picture
is of one low-mass $\W$, which decays very rapidly, but with decay
products swiftly moving away from the central region, while the
heavier $\W$ is essentially on the mass shell and therefore decays
later. The decrease in ${\cal P}_{\mrm{recon}}$ below 160~GeV thus
reflects that the lighter $\W$ has fewer decay products, and that
those products are faster boosted away from the origin.

The variation of ${\cal P}_{\mrm{recon}}$ with c.m. energy is a bit
more pronounced in scenario II than in scenario I. This is fully
understandable, in the sense that a faster motion of the
$\W$'s away from each other implies that the $\W^+$ and $\W^-$ decay
products are better separated. The probability for a head-on crossing
of two string cores, as required in scenario II, therefore dies away
faster with increasing $p^*$ and $\beta$ than the probability for the
Gaussian tails of the strings to overlap, as is sufficient in
scenario I.

The difference between the two scenarios is even more marked in
Fig. 15b, where the dependence of ${\cal P}_{\mrm{recon}}$ on the parton
shower cut-off scale is shown for a fixed energy $E_{\mrm{cm}} = 170$ GeV.
The $k_{\mrm{I}} = 0.6$ choice is made such that the two curves
agree for the nominal cut-off $m_0 = 1$ GeV. In the other limit,
when $m_0 \geq \mW/2$, there is no parton shower evolution, i.e.
each $\W$ gives a simple $\q\qbar$ system.

When each $\W$ only decays to one string piece, the solution to eq.
(\ref{stringcross}) in scenario II is $t=0$,
$\alpha^+ = \alpha^- = 0.5$, i.e. in the unphysical region before
the $\W$'s have decayed. In other words, if there is no parton
shower evolution, the two string pieces are moving away from each
other and cannot cross. The reconnection probability thus has to
vanish in the large $m_0$ limit.
Conversely, the more string pieces there are, the larger is the
probability that two of these pieces will cross. However, the overall
string drawing is mainly determined by the energetic gluons emitted at
large virtualities, while low-virtuality branchings only add smaller
wrinkles to this. ${\cal P}_{\mrm{recon}}$ therefore does not increase as
fast as the product of the numbers of $\W^+$ and $\W^-$ string pieces.

By contrast, scenario I only shows a small dependence on the parton
shower cut-off scale: the $\sim 0.5$ fm thick strings produced
$\sim 0.1$ fm apart can overlap even if no gluons at all are emitted,
and introducing further structure to these strings can go both ways,
with only a small net increase in overlap. To understand this,
compare a $\q\qbar$ and a $\q\qbar\g$ decay of a $\W$ at rest. In the
former, the string is expanding out on two sides but always passes
through the origin. In the latter, both string pieces
($\q\g$ and $\g\qbar$) are
boosted away from the origin. Two $\W$'s that decay to multijets
will therefore usually have a smaller overlap close to the
middle of the event than would two $\W$'s decaying to $\q\qbar$,
which partly compensates for the gain in overlap away from the middle.

In Section 4.2 we argued that shower evolution at scales above 10 GeV
occur at distances smaller than the separation between $\W^+$ and
$\W^-$ decays. However, also a parton with a virtuality below 10 GeV
would often have time to branch and produce a new string piece which
could interact --- the $\W^+$--$\W^-$ separation only sets a lower
limit for how soon strings meeting head-on can collide. Further,
because of the way the shower algorithm is constructed, the usage of
an $m_0 = 10$ GeV cut-off in fact also removes some branching
configurations that would be allowed for an $m=10$ GeV parton in a
shower with $m_0 = 1$~GeV. While it may be that a shower with
$m_0 = 1$ GeV gives a string with more wrinkles than is realistically
resolved at the time of overlap, the above arguments indicate that an
$m_0 = 10$~GeV scenario is too far in the other extreme. A realistic
range of uncertainty for ${\cal P}_{\mrm{recon}}$ would probably correspond
to the variation between $m_0 = 1$~GeV and $m_0 = 2$--3~GeV, i.e. at
most a factor of 2 for scenario~II and negligibly little for
scenario~I.

Even for c.m. energy, $m_0$ cut-off and other parameters fixed,
the reconnection probability varies from one event to the next as a
function of the partonic configuration generated on the perturbative
level. Generally speaking, events with large partonic activity also
have a large ${\cal P}_{\mrm{recon}}$. Fig. 16a shows the dependence of
${\cal P}_{\mrm{recon}}$ on the number of reconstructed jets (using
{\tt LUCLUS} with the default $d_{join} = 2.5$ GeV \cite{Jetset}).
The reconnection probability varies by about a factor of 2 between
small and large jet multiplicities. As explained in the previous
paragraphs, results in scenario II are more sensitive to the
number of partons, and therefore show a larger variation of
${\cal P}_{\mrm{recon}}$ with $n_{\mrm{jet}}$. Since a larger number
of jets also translates into a larger charged multiplicity,
${\cal P}_{\mrm{recon}}$ varies almost as much with
$n_{\mrm{ch}}$ as with $n_{\mrm{jet}}$, Fig. 16b.
Again differences between scenarios I and II
are obvious. Among simpler measures, the charged multiplicity seems
to be the one that gives the largest variation in
${\cal P}_{\mrm{recon}}$. By contrast, global event measures such
as thrust mainly reflect the topology of the
$\q_1\qbar_2\q_3\qbar_4$ configuration (before parton showers) and
${\cal P}_{\mrm{recon}}$ is therefore rather insensitive to them.
The $T_4$, i.e. the thrust minimized with respect to four different
jet directions, is of about comparable quality with $n_{\mrm{ch}}$.

\subsection{Event properties}

In Fig. 2 we showed how large the effects of reconnections can be in
the case of a particularly favourable event topology and instantaneous
reconnections. It is now time to see how much of that survives. We
have already demonstrated that the instantaneous reconnection scenario
is incompatible with perturbative QCD calculations. Furthermore, phase
space is vanishing for the topology of Fig. 2, i.e. with the $\W^+$
and $\W^-$ at rest. While some selection might still be made based
on event topology, to first approximation it is more realistic to
integrate over all configurations at a fixed c.m. energy. The
intermediate reconnection scenario shown in Fig. 3 is therefore a
more reliable guide to how large are the effects to be expected.
With the
exception of the $\W$ mass plot, differences to the no-reconnection
baseline scenario are small. The results of Fig. 3 were still for
100\% reconnected events. A realistic scenario with $\sim$30\%
reconnection, as obtained above, would therefore be expected to show
even smaller differences to the no-reconnection scenario. The
reduction would not have to be exactly by a factor of 0.3, since our
scenarios I and II allow reconnections to take place between any
two string pieces, while the intermediate scenario assumes all
reconnections to occur in the middle of the event.

As expected, the charged multiplicity distribution comes out very
similar with and without reconnection, Fig. 17a. The averages are
$\langle n_{\mrm{ch}} \rangle =$
36.63, 36.31 and 36.45 for no reconnection, scenario I and
scenario II, respectively. These numbers are based on a Monte Carlo
sample of 40,000 events, so with a width of the multiplicity
distributions of about 7.9 this gives a statistical error of 0.04.
The differences between the models thus are statistically
significant, but are at a level well below
experimental observability: LEP 2 will have to face lower statistics,
a large background from other processes, QED radiation from the
initial and final state, and a need to construct the expected
no-reconnection multiplicity by extrapolation from $\Z^0$ data at
LEP 1. On top of that, essentially all the multiplicity difference
is appearing in the low-momentum region, $|\mbf{p}| < 1$ GeV, where
detector efficiency may be limited.

Unfortunately, effects seem to be very small also for the other
simple event properties we have studied, such as the rapidity
distribution $\d n_{\mrm{ch}} / \d y$, the number of charged particles
in $|y|<1$, the thrust and $T_4$ distributions, and the number of
jets. Also trivial correlations, such as the average charged
multiplicity as a function of the number of jets, show no significant
effects.

It will therefore be necessary to devise more specific measures,
which are intended to probe in detail the regions of phase space where
differences are expected. This will be a delicate task, and we have no
good proposal at the moment. As an example, which again does not show
sufficiently large effects to be interesting in itself, consider
Fig. 17b. Exactly four jets are found in each event with {\tt LUCLUS},
and all particles are assigned to one of these jets. The azimuthal
distribution of particles around the jet axis is plotted for
particles belonging to the jet, with $\varphi = 0$ defined by the
nearest other jet. No attempt is made to specify an orientation around
the jet axis to distinguish $-\varphi$ from $+\varphi$, so the range
is $0 \leq \varphi \leq \pi$. For the plot in Fig. 17b we have
additionally required that the opening angle to the nearest jet
should be at least 1 radian, and only considered particles with a
transverse momentum relative to the jet axis in the range
0.3 GeV $< p_{\perp} <$ 0.8 GeV. These cuts slightly enhance the
dynamical effects discussed below, but are not essential.

To zeroth approximation one would expect a flat distribution
$\d n / \d \varphi$. To first approximation, the particle
density should be suppressed close to $\varphi = 0$, since
particles which emerge in that direction can easily be assigned
to the nearest other jet instead. There should also be a somewhat
smaller suppression close to $\varphi = \pi$, around the
next-to-nearest other jet --- remember that the event
topology is essentially that of two back-to-back jet pairs,
one pair from each $\W$, so the nearest and next-to-nearest jets
should be close to opposite in azimuth. On top of this now comes
potential dynamical string effects. The original string stretchings
are between almost back-to-back jets, and therefore are not expected
to give large azimuthal anisotropies. In addition, with our choice of
$\varphi = 0$, the azimuthal angle of the
almost back-to-back jet is fairly isotropically distributed.
If a reconnection occurs, it will therefore show up as a string
stretched either to the nearest or to the next-to-nearest other jet,
i.e. lead to more particle production close to $\varphi = 0$ or
$\varphi = \pi$ and less at around $\varphi = \pi/2$. Unfortunately,
it seems that effects in scenarios I and II are too small to be
detectable, although the results for the (unphysical) instantaneous
scenario confirm that the distribution is in principle sensitive to
effects.

A more ambitious goal would be to search for phenomena that would
be impossible without reconnections. However, the ones we have been
able to think of are all flawed. Consider e.g. the production of
particles at large $x = 2E/E_{\mrm{cm}}$. The kinematical limit is
$x \leq 1$ in an $\ee \to \gamma^*/\Z^0 \to \q\qbar$ event, but less
in an $\ee \to \W^+\W^- \to \q_1\qbar_2\q_3\qbar_4$ one, because the
energy is shared between more primary partons. For instance, for the
two $\W$'s at rest and with equal masses the limit is $x \leq 1/2$.
So an excess at large $x$ would be a signal that particles are
allowed to pick up energy from both the $\W^+$ and $\W^-$ during
the fragmentation stage, i.e. evidence for reconnections. However,
the string topology has to be very special for large-$x$ production
to be possible at all, with e.g. the $\q_1$ and $\qbar_4$ closely
parallel. Therefore we do not observe a large-$x$ excess in the
reconnection scenarios. Even if we had done so, the excess would have
had to be very large to be able to compete against the
background of $\ee \to \gamma^*/\Z^0 \to \q\qbar\g$ events with the
same topology, where large-$x$ production is not suppressed.

String reconnection would also allow a new way of producing $\JP$
mesons: in a $\W^+\W^- \to \c\sbar \, \s\cbar$ event the $\c$ and
$\cbar$ may be reconnected into a colour singlet, which `collapses'
into one single $\JP$ particle. (If the $\c\cbar$ invariant mass
is too large, the system will fragment into a pair of $\D$ mesons
plus further hadrons, a final state which is indistinguishable
from no-reconnection $\D$ meson production.) In a sample of 50,000
scenario~II hadronic $\W^+\W^-$ events we found one such occurrence.
This is overwhelmed by a ten times larger background from
$\c \to \c \, \g \to \c \, \cbar \c \to \JP \, \c$,
i.e. parton shower branchings followed by a `collapse' of a low-mass
colour singlet $\c\cbar$ system, and $\W^+ \to \c\bbar$
decays and $\g \to \b\bbar$ shower branchings with a $\B$ hadron
decaying to $\JP$. In an experiment with several hundred times the
expected LEP 2 statistics, one could presumably also require a large
$x$ to enhance the signal, but the background from processes such as
$\gamma^*/\Z^0 \to \b\bbar\g$ would almost certainly be overwhelming.

We close this subsection with an indirect investigation of string
reconnection effects, viewed as a function of the position of the
string piece where the reconnection is occurring. In our approach it
is not meaningful to try to define position better than this,
i.e. to a specific point along the string piece, since the whole
piece is affected by a reconnection. If partons are ordered along
the (original) string from one end to the other, and partons on
the two sides of the affected string piece have total energies
$E_1$ and $E_2$ in the rest frame of the $\W$, then the
reconnection position can be defined by
\begin{equation}
d_{\mrm{rec}} = \frac{|E_1 - E_2|}{E_1 + E_2} ~,
\end{equation}
where the denominator is nothing but the mass of the $\W$.
A $d_{\mrm{rec}} \approx 0$ corresponds to a reconnection close to the
middle of the string, while a $d_{\mrm{rec}} \approx 1$ is a
reconnection close to the string endpoint. Each reconnection
is characterized by one $d_{\mrm{rec}}$ for the $\W^+$ and one for the
$\W^-$; below we only consider the average
$\overline{d}_{\mrm{rec}} = (d_{\mrm{rec}}^- + d_{\mrm{rec}}^+)/2$.
The distribution of $\overline{d}_{\mrm{rec}}$ values is shown in
Fig. 18a. The mean is
$\langle \overline{d}_{\mrm{rec}} \rangle \approx 0.18$.
However, this number comes from the convolution of two effects,
the one quoted above and the parton shower evolution. The latter
allows the two jets to have different masses and thus
$E_1 \neq E_2$. In the intermediate scenario, where na\"{\i}vely
reconnections occur at $\overline{d}_{\mrm{rec}} \equiv 0$, parton
showers actually give a
$\langle \overline{d}_{\mrm{rec}} \rangle \approx 0.06$.

One would like to study how a reconnection changes e.g. the
multiplicity
of an event, as a function of $\overline{d}_{\mrm{rec}}$. However,
effects are too small to be studied well that way, given that
fluctuations in the fragmentation process are non-negligible.
Instead we use the $\lambda$-measure of string length
\cite{lambda}, which is roughly proportional to the expected
hadronic multiplicity but does not contain the fragmentation
fluctuations. The complete formulae for $\lambda$ are quite
complicated, but it is here enough to use the simpler form
\begin{equation}
\lambda = \sum_i \, \ln \left( 1 + \frac{m_i^2}{m_{\rho}^2}
\right) ~.
\end{equation}
The sum runs over all string pieces $i$ in the event, with
$m_i$ the string piece invariant mass (calculated from the
momenta of the partons that span the string, with gluon momenta
split equally between the two adjacent pieces), while $m_{\rho}$
sets a suitable hadronic mass scale. A reconnection means that two
terms in the sum have to be replaced by two new terms. The
change induced by a reconnection is defined as
$\Delta \lambda = \lambda_{\mrm{new}} - \lambda_{\mrm{old}}$.

Fig. 18b shows how $\langle \Delta \lambda \rangle$ varies as a
function of $\overline{d}_{\mrm{rec}}$. To set the scale,
$\langle \lambda \rangle \approx 20$. In fact, not much variation is
observed in scenario I. In scenario II there is a clear trend
that reconnections at larger $\overline{d}_{\mrm{rec}}$ give a smaller
change $\Delta \lambda$. In part, this is related to a larger
admixture of reconnections with a topology like the one shown in
Fig. 14c. In scenario II$'$, where events with an increased string
length are removed, almost the same behaviour is recovered as in
scenario I. (The cut in scenario II$'$ is in terms of the
criterion $m_{14}m_{23} > m_{12}m_{34}$, which is essentially
equivalent to $\Delta \lambda > 0$.) This does not fully explain
why the curves are so flat for scenario I, since also this scenario
contains reconnections with $\Delta \lambda < 0$, although fewer,
and these were not removed.

In principle, the $\overline{d}_{\mrm{rec}}$ and $\Delta \lambda$
quantities could be related to experimental observables, such as
the event topology and multiplicities in different regions of the
event. Unfortunately, the results above then indicate that
it may be difficult to distinguish between scenarios I and II
experimentally, given that minor variations of scenario II may
give a spread in observables as big as the one between
scenarios I and II in the first
place. Further studies are therefore needed to understand the
implications of reconnections as a function of event topology
and reconnection scenario, and to invent more sensitive
experimental observables.

\subsection{W mass determinations}

The most critical single observable for LEP 2 physics is the $\W$
mass. Experimentally, $\mW$ depends on all particle momenta of an
event in a non-trivial fashion.

If there are no interference effects between the $\W^+$ and $\W^-$
decays, each final state particle uniquely comes from one of the
decays. Given a correct separation of particles into these two
classes, the $\W^{\pm}$ four-momenta can be reconstructed and
squared to give the $\W^{\pm}$ masses. In practice, several
complications have to be faced:
\begin{Enumerate}
\item the purely statistical error is $\sim \GW / \sqrt{n}$ for
$n$ observed $\W$'s, from the fluctuations in the mass of produced
$\W$'s;
\item the average $\W$ mass at LEP 2 is not the fundamental $\W$
mass parameter of the theory, since it is offset because of
phase-space and (Born-level) matrix-element distortions,
and of higher-order QED and weak corrections;
\item background events may be mistaken for signal ones;
\item particle four-momentum determinations have experimental
errors;
\item some particles are not observed at all, e.g. neutrinos from
charm decays and particles at low angles to the beam direction;
\item initial state photons may go unobserved or may
(mistakenly) be assigned to either $\W$;
\item a particle coming from the $\W^+$ decay may be assigned to the
$\W^-$, and vice versa; and, finally,
\item QCD interference effects, perturbative or non-perturbative,
may efface the separate identities of the $\W^+$ and $\W^-$ systems.
\end{Enumerate}

It is not our intention to cover points 1--6 here; for a study
of those aspects see ref. \cite{LEP2work}. The first two points are
sidestepped by
comparing, event by event, the actually generated $\W$ masses with the
reconstructed ones. (For an 80 GeV input mass the actually generated
average mass is 79.2 GeV at our standard energy of 170 GeV. The
0.8 GeV difference, or whatever a more detailed formalism gives,
can always be compensated analytically at the end of the day.)
Background processes are not at all studied; presumably they can be
controlled by fairly standard cuts. Points 4--6 would experimentally
be addressed by a variety of techniques, including making use of
energy--momentum constraints, based on the known beam energy.
We, however, will assume that all particles are perfectly well
measured.

If a particle or a collection of particles (of reasonably small
total energy) are misassigned between
the $\W^+$ and $\W^-$, it is easy to see that the $\W^+$ and $\W^-$
mass estimates will be shifted in opposite directions. The shifts
do not have to have the same size, but a significant cancellation
of errors is obtained if the average
$\overmW = (\mW^+ + \mW^-)/2$ is used rather than the
individual $\W$ masses. The cancellation more than
compensates for the statistics loss of having only one $\W$ mass
number from each event rather than two. Experimentally, the
assumption that the two $\W$ masses of the events are the same
is even more useful: if the beam energy is known and initial state
QED radiation is neglected then only one number need be determined,
namely $p^* = |\mbf{p}^+| = |\mbf{p}^-|$. This number mainly depends
on the (acollinearity) angles between jets, which presumably can
be measured well, while less well measured jet energies can be
constrained by kinematics.

To study the effects of point 8, and those of point 7 that survive
the $\W$ mass averaging, the following procedure is adopted. The
{\tt LUCLUS} algorithm \cite{Jetset} is used to reconstruct four
or more jets, with the jet distance parameter $d_{join} = 8$ GeV.
This gives almost 25\% of events with five or more jets. These
events are eliminated from the subsequent analysis, since events
which contain hard gluon radiation give much worse $\W$ mass
resolution --- a whole jet may be misassigned. Further, it is
required that each jet has an energy of at least 20 GeV and that
the distance between any two jets is larger than 0.5 radians ---
misassignment of particles is much more likely if two jets are
nearby. In total about 60\% of the events survive all the cuts
(the number goes up to about 64\% for the instantaneous scenario,
owing to a reduced amount of QCD radiation). The cuts have not been
optimized, but should constitute a sensible first approximation.
For the surviving events, the jet--jet invariant masses are
calculated.

The four jets may be paired in three different ways, so that
each event yields three possible $\overmW$ values.
Very often one combination stands out as the most likely one,
but ambiguities are not so infrequent. It is quite feasible to
histogram all three combinations and treat the wrong ones as a
background to the peak. We have instead tried a few different
criteria to pick one of the three combinations.
\begin{Enumerate}
\item In a Monte Carlo one has the advantage of knowing the
original $\q_1\qbar_2\q_3\qbar_4$ configuration, before parton
showers and fragmentation. The reconstructed jets are thus
matched one-to-one to the original partons; this is done by
picking the matching that minimizes the product of the
invariant masses between each jet and its associated parton.
The subdivision into $\W^+$ and $\W^-$ jets is thereafter
automatic.
\item Since we already know that the $\W$ mass is close to
80 GeV, one may pick the combination that minimizes
$|\overmW -80|$.
\item It could be argued that the above recipe would accept
false combinations, where one mass is too small and the
other too large, so alternatively one could instead minimize
$|\mW^{(1)} -80| + |\mW^{(2)} -80|$, where $\mW^{(1,2)}$ are the
two reconstructed jet--jet masses.
\item Close to threshold the jets from the same $\W$ should
be almost back-to-back. One can therefore maximize the sum of
opening angles between jets from the same pair.
\end{Enumerate}

By and large, these procedures give comparable results.
In particular, the use of parton-level information in the first
procedure does not offer significant gains --- ambiguities
in events mainly arise because of hard QCD radiation, and then
the original $\q_1\qbar_2\q_3\qbar_4$ configuration is not
too reliable a guide.

The results presented in Section 2 and Fig. 3d are based on
the first method. Corresponding results for scenarios I and
II are shown in Fig. 19a, while Figs. 19b--19d show results
for methods 2, 3 and 4. Considering the no-reconnection events,
method number 2 is the one that gives the smallest systematic
offset, $\langle \Delta \overmW \rangle =
\langle (\overmW)_{\mrm{reconstructed}} -
(\overmW)_{\mrm{generated}} \rangle$, see Table 1.
Method number 1 gives the narrowest distribution, defined in terms
of the width of the $\Delta \overmW$ distribution in the
range $\pm$10 GeV. Since this method makes use of knowledge
hidden in real life, unlike the others, it is surprising that the
advantage is not bigger than it is. Method number 4, which makes
no assumption about the $\W$ mass value, is the least successful
one. The tails of the $\Delta \overmW$ distributions are very
broad and act to push up the width. To the spreads
$\sigma(\overmW)$ of Table 1 should
be added the effect of the far wings, when $|\Delta \overmW|$
is larger than 10 GeV. Typically 2\%--4\% are found in these regions;
predominantly $\Delta \overmW<10$ GeV for methods 1 and 3 and
$\Delta \overmW>10$ GeV for methods 2 and 4. The importance of
the wings should not be overstated; any actual fitting program
for the $\W$ mass would be mostly sensitive to events in the
region close to the peak.

The systematic mass shifts obtained for the no-reconnection
scenario are not so critical: they can be calculated by standard
Monte Carlo methods and corrected for. Any method that performs
well in other respects can therefore be adopted, even when it is
known to induce a non-negligible systematic shift. Worse is any
additional shift appearing when reconnection effects are considered.
Such shifts may have to be included among the systematic errors
of an experimental $\W$ mass determination. Also other aspects of
the distributions could be of some importance, such as the
differences in peak height visible in Fig. 19.

The systematic mass shifts of scenarios I and II are shown in
Table 1. Numbers are also given for the intermediate and
instantaneous scenarios: although we know these latter two to be
unphysical, they are somewhat easier to study because of the
larger size of the shifts. If shifts are suitably scaled down, the
instantaneous scenario may also give some feel for the size of
potential perturbative effects. Rather surprisingly, the sign
of the mass shifts need not be the same. Specifically, while
the reconstructed $\overmW$ always is shifted upwards in the
intermediate and instantaneous scenarios, compared to the
no-reconnection one, the shifts are very small and of uncertain
sign for scenario I and large and negative for scenario II.

We have no detailed understanding of these mass shifts, but a
few general comments can be made. One way of viewing the
$\W$ mass determination is to consider the angles between the
four jets. If those jets appear as two back-to-back pairs then
the event is reconstructed as two $\W$'s at rest, with a mass sum
equal to the c.m. energy. The further away from 180$^{\circ}$
the angle between the two jets of a $\W$ gets to be, the lower is
its mass, other things being the same. Now consider the two original
massless partons from one of the $\W$ decays, where the
$\W$ is not at rest. The $\W$ mass is then given by
\begin{equation}
m^2 = (p_1 + p_2)^2 = 2E_1 E_2 (1 - \cos\theta_{12})
{}~.
\label{mrecsimp}
\end{equation}
The fragmentation products are boosted in the $\W$ direction of
motion, and thus soft particles are shifted into the angular
range between partons 1 and 2 (the familiar string/dipole result).
This leads to a systematic shift inwards of reconstructed jet
directions \cite{JADE}, i.e. a reduction of $\theta_{12}$.
Were eq. (\ref{mrecsimp}) to be used, this would lead to an
underestimate of $m$. However, the fragmentation also implies
that the two reconstructed jets acquire masses $m_{1,2}$,
so that the correct formula to use is
\begin{equation}
m^2 = m_1^2 + m_2^2 + 2 E_1 E_2 - 2 \sqrt{E_1^2 - m_1^2}
\sqrt{E_2^2 - m_2^2} \cos\theta_{12} ~.
\label{mreccomp}
\end{equation}
The introduction of jet masses exactly compensates the reduction
in $\theta_{12}$.

Let us now assume that a reconnection occurs before the
fragmentation. Depending on how the new strings are stretched,
i.e. how soft particles are distributed between jets, the
reconstructed $\theta_{12}$ may be shifted in different directions.
As a first approximation, imagine that this shift is completely at
random, i.e. that the new colour partner is isotropically
distributed in azimuth with respect to the old partner.
On the average, $\theta_{12}$ then remains unchanged in the
fragmentation stage, rather than being reduced. If eq.
(\ref{mreccomp}) is still used,
the mass terms will then lead to $m$ being overestimated.
This is likely to be what is going on in the intermediate and
instantaneous reconnection scenarios.

In scenarios I and II, reconnections do not necessarily occur
close to the middle of events, i.e. close to the original $\W^+\W^-$
production and decay volume. Many low-momentum particles will
therefore be distributed along the na\"{\i}ve string direction,
as if no reconnection had occurred. The mass shift then also comes
from other regions of the event, which are less transparent
to analyse. Branchings at the end of the parton showers (which
occur late and therefore should maybe not be included in the
reconnection description) do not seem to play any particular
r\^ole: results with $m_0 = 2$~GeV are consistent with the
default $m_0 = 1$ GeV ones. Also note that the results for
scenario~II$'$ are very close to those of scenario II, i.e. here
reconnections that increase the string length do not seem to
have any special significance, unlike the results of Fig. 18b.
It is tempting to speculate that the observed difference could be
related to the fact that string crossings have to be `head-on'
in scenarios II and II$'$ while they need only be grazing in
scenario I, and that this translates into different sets of events
for which reconnections occur away from the middle.

The numbers in Table 1 are based on a total statistics of
160,000 events for each of the scenarios considered,
whereof $n \approx 100,000$ survive the cuts. (The statistical
error on the $\W$ mass will be limited by the number of LEP 2
events. Here we are considering the statistical error on an
evaluation of a systematic error, and for this purpose
Monte Carlo statistics well in excess of the experimental
sample is not inappropriate.) This gives a typical
error of $\sigma(\overmW)/\sqrt{n} \approx 2~\mrm{GeV}/10^{2.5}
\approx 6$~MeV for each $\langle \Delta \overmW \rangle$
determination in a specific scenario and a factor of $\sqrt{2}$
larger than that for the systematic mass shift between scenarios.
While systematic effects in the intermediate and instantaneous
scenarios are thus established beyond doubt, the size of effects
in scenarios I, II and II$'$ remains ill defined. From Table 1, one
could assume that non-perturbative effects are bounded by 30 MeV,
say. Admittedly, some numbers for scenarios II and II$'$ are
slightly larger than this, but optimized methods are likely
to do somewhat better.
Although we primarily base ourselves on scenarios I and
II, the 30 MeV number is also consistent with the intermediate
scenario scaled down by ${\cal P}_{\mrm{recon}} \approx 1/3$.
We have argued that perturbative effects are strongly
suppressed, by at least a factor of $10^{-2}$ compared with the
instantaneous scenario, so any perturbative systematic effects
should at most be at the 5 MeV level. There could also be
an interplay between the perturbative and non-perturbative
phases (see below in the summary). We have no way of estimating
the size of such effects, but would not expect them to be
larger than the ones we already studied. So assume it to be another
5 MeV, with numbers to be added linearly since they are not
unrelated.

In total, our best estimate of the systematic uncertainty on
the $\W$ mass would thus be a number roughly like 40 MeV from
the reconnection phenomenon alone. In view of the limit set
by LEP 2 statistics, this number is manageable, but not
negligible. It may well be the largest individual source of
systematic error for $\W$ mass determinations in doubly
hadronic $\W^+\W^-$ decays \cite{LEP2work}. However, note that
our $\W$ mass reconstruction methods are not the experimental
ones. Other methods may be more or less sensitive to the
differences between the scenarios: today we do not know.
It is therefore important that each experimental $\mW$
reconstruction method is checked against the scenarios in
this paper. For instance, the importance given to the wings of the
mass peak could quite likely be tuned to reduce the uncertainty
by some amount. It is also important that the data are checked for
any trace of reconnection effects. So far we did not find the
good variable(s), but if the $\W$ mass is shifted by reconnection
effects, this must mean that some
particles somewhere in the event are affected. Detailed
experimental studies, based on a deepened understanding of the
reconnection phenomenon, should therefore allow a self-screening
of the data to be carried out, with the possibility to apply
systematic mass corrections as a function of the observed
deviations from the no-reconnection scenario.

\section{Summary and Outlook}

It should by now be clear that the phenomenon of colour
reconnection --- which has to be there, at some level (remember
$\B \to \JP$) --- is not totally understood. In this paper we have
shown why perturbative colour rearrangements are strongly
suppressed, while non-perturbative recouplings may be quite
frequent. However, we are aware that the paper may raise
more new questions than it answers old ones.

The main points of the pertubative description
of the $\W^+\W^- \to \q_1\qbar_2\q_3\qbar_4$ production not
far from threshold are the following:
\begin{Itemize}
\item Any reconnection effects come with a factor
$\alpha_s^2/(N_C^2-1)$, i.e. they cannot appear in first but
only in second order, and are colour-suppressed.
\item The difference between the $\W^+$ and $\W^-$ decay times
implies a relative phase between the radiation accompanying
the $\W^{\pm}$ decays. This leads to a suppression of the
interference terms for gluon energies $\omega \gtrsim \GW$.
Interference effects are thus limited to a few low-momentum hadrons.
\item In a $\W^+\W^- \to \q_1\qbar_2\q_3\qbar_4$ decay the original
dipole structure is $\widehat{12} + \widehat{34}$, while colour
transmutation corresponds to an admixture of the further dipole
combination
\mbox{$\widehat{14} + \widehat{23} - \widehat{13} - \widehat{24}$}.
The appearance of negative-sign dipoles can only be understood
in an inclusive sense, i.e. the complete dipole combination is
active in the production of particles in each hadronic event.
\item In the total cross section, the cancellation between real
and virtual soft gluon emission leads to the interference terms
being suppressed by at least a factor of $\GW/\mW$, in addition
to the overall factor $\alpha_s^2/(N_C^2-1)$ noted above.
\end{Itemize}
Of course, the degree of suppression of interference terms needs
to be analysed for each specific observable. The suppression need
not always be large. Generically, the higher the degree of
inclusiveness, the stronger the suppression.

In order to address non-perturbative colour reconnection effects,
we have developed several alternative models for the space--time
structure of the string fragmentation process. To the best of our
knowledge, nothing similar has ever been attempted before.
Our two main scenarios are called I and II, by analogy with
type I and type II superconductors, respectively. The type II
superconductor is often used as a toy model for confinement in QCD,
while scenario I is closely related to the bag model approach to
QCD wave functions, so both have some credibility.
Nature could well correspond to something in between
these extremes, however, or even to something altogether different.
Unlike the (amplitude-level-based) perturbative
description, the (probabilistic) non-perturbative string always
corresponds to a well-defined colour topology of the event: you
either have strings $\q_1\qbar_2$ and $\q_3\qbar_4$ or strings
$\q_1\qbar_4$ and $\q_3\qbar_2$, not both at the same time. And
you do not have the negative-sign dipoles $\q_1\q_3$ and
$\qbar_2\qbar_4$.

To make any progress at all, we have had to rely on models that
are far from perfect. The uncertainty comes in many forms:
\begin{Enumerate}
\item As a positive asset:
the fact that different models for the colour string give
different predictions means that it may be possible to learn
about the structure of the QCD vacuum. It is here essential
that the colour rearrangement is almost exclusively a
non-perturbative phenomenon. Had perturbative aspects dominated,
the whole topic would just have been one of testing our
calculational skills, without any chance to learn something
fundamentally new.
\item Because of `laziness': we have introduced various
approximations in the description of the string motion and the
fragmentation process. With a lot of hard work, but without the
need for any truly new ideas, it would be possible to improve on
these aspects. The amount of work would be out of proportion with the
potential gains in precision, however, so long as there is no
possibility to receive any experimental feedback.
\item As a true limit in our current physics understanding:
one example is the question of an interplay between the
perturbative and non-perturbative phases, see discussion below.
\end{Enumerate}

To begin with, consider the production of two $\W$'s at rest.
One of the two will decay first. The partons
of this decay, both the original $\q\qbar$ ones and additional
bremsstrahlung gluons (and $\q'\qbar'$ pairs), will spread outwards
from the origin with the speed of light. The partonic activity is
therefore concentrated in an outward-moving shell of roughly
$1/\mW$ thickness, if one just considers the $\W$ decay process.
The production of massive partons in the shower evolution increases
the width of the shell by some amount, but does not change the
overall picture. This shell is not equally populated,
but contains `hot spots' around the directions of the outgoing
$\q$ and $\qbar$, while most of the shell is empty. Inside the
expanding shell there are no perturbative partons, but only the
softer confinement gluons.

When the second $\W$ decays, there is no possibility for its
perturbative partons to interact directly with the pertubative
partons of the first $\W$: the first shell is always spreading
ahead of the second one. However, the perturbative partons
of the second $\W$ may directly interact with the left-behind
confinement partons (i.e.\ the strings or colour flux tubes) of
the first. This phenomenon is what we mean  by an interplay
between the perturbative and non-perturbative phases. Maybe it
could one day be addressed as `$\W$ decay and parton shower
evolution in a heat bath', with techniques similar to those
being developed in heavy-ion physics. For now, it is completely
neglected. All colour rearrangements are therefore attributed to
the interactions between the confinement gluons of the two $\W$
decays.

One can obtain very large reconnection effects with optimistic
physics assumptions applied to idealized experimental conditions,
cf. the GPZ study. However, if we consider a realistic physics
model in a real-life situation, like $\W^+\W^-$ decay at LEP 2,
all studied effects turn out to be very small. The most spectacular
one is the implied 40 MeV systematic uncertainty on the $\W$
mass. The reader may well wonder: how come it is so difficult to
find other variables that show large effects, when we already
have one? Well, in fact, the $\W$ mass shifts we
discuss here are not large, it is only that we are set to measure
$\mW$ with the highest possible accuracy. A 40 MeV uncertainty
corresponds to a relative error on $\mW$ of half a per mille.
We already noted that the total charged multiplicity shows
effects a factor of 10 bigger than this --- but also concluded
that various uncertainties will make $\langle n_{\mrm{ch}} \rangle$
useless as a probe of reconnection effects. Maybe the best we can
hope for is a probe that gives differences, if only at the per cent
level, but where the no-reconnection prediction is very well
understood. In the long run, we believe such measures will be
constructed.

One interesting cross-check on models could potentially be provided
by Bose--Einstein effects. The typical separation of
$\W^+$ and $\W^-$ decay vertices is smaller than the observed
Bose--Einstein radius, so na\"{\i}vely all hadrons could be viewed
as produced from a single source. However, if the strings from
the $\W^+$ and $\W^-$ decays keep their separate identities,
i.e. act as independent sources,
the observable magnitude of Bose--Einstein effects will be
smaller in $\W^+\W^-$ events than in $\Z^0$ ones. A behaviour
of this kind may already have been seen in the UA1 data
\cite{UA1}. It would be particularly exciting if the size of
Bose--Einstein effects at LEP 2 varied as a function of event
topology in a way related to the reconnection probability.

Our current 40 MeV estimate for systematic uncertainties on the
$\W$ mass is the sum of 30 MeV from non-perturbative reconnections,
5 MeV from perturbative ones and (out of thin air) 5 MeV from
the interplay of the perturbative and non-perturbative phases.
The 30 MeV number depends sensitively on the assumed reconnection
probability, which is a completely free parameter in scenario I
and has a non-negligible uncertainty in scenario II. (Most notably,
maybe strings can only reconnect the third of the time
when the endpoint colours match.) However, the reconnection
probability cannot be too low, or one would not observe $\JP$
in $\B$ decay. Even if, arbitrarily, one were to reduce the
non-pertubative contribution by a factor of 2,
the net uncertainty would still be 25 MeV, which is not quite
negligible. The issue therefore has to be studied further, e.g.
for the choice of the best experimental algorithm to use.

Despite its evident shortcomings (large missing momentum) the
mixed leptonic--hadronic channel is free from the potential
reconnection-induced ambiguities. Also QED-induced interference
effects would be under control \cite{K4,K5,K6}. The relative
importance to be assigned to results from different $\W^+\W^-$
decay channels therefore has to be assessed carefully.

A high-statistics run above the $\Z^0\Z^0$ threshold would allow
an unambiguous determination of any systematic mass shift,
given that the $\Z^0$ mass is already known from LEP~1 \cite{Inno}.
Such a shift could then be used to correct the $\W$ mass.
If the various potential sources of systematic error could be
disentangled, it could also imply a direct observation of
reconnection effects. More generally, $\Z^0$ events from LEP 1
can be used to predict a number of properties for $\Z^0\Z^0$
events, such as the charged multiplicity distribution.
Any sign of deviations would provide important information on
the reconnection issue.

Perturbative and non-perturbative calculations alike show only a
very slow variation of effects with c.m. energy: there is no
essential variation over the range of $\W^+\W^-$ energies that
can be probed by LEP 2. If we want to call colour reconnection a
threshold effect, we have to acknowledge that the threshold region
is very extended.

In some sense, the effects studied here could be considered as only
the tip of an iceberg. Colour reconnection can occur in any process
which involves the simultaneous presence of more than one colour
singlet, such as $\t\tbar$ events, decays of colour singlet resonances
produced in hadron colliders, and so on. Many of the techniques
developed for $\W^+\W^-$ could be directly applied to these problems.
Additional aspects may have to be addresses, such as the presence of
colour in the initial state for hadron colliders.

A related set of phenomena can occur in reactions with only one
colour singlet, and hence without the possibility of a colour
rearrangement in the sense of this paper. Specifically, we have
in mind a simple process such as
$\e^+\e^- \to \t\tbar \to \b\W^+ \, \bbar \W^- \to
\b \ell^+ \nu_{\ell} \, \bbar \ell^- \overline{\nu}_{\ell}$.
If the top mass is sufficiently large ($m_{\t} \gtrsim 120$~GeV)
and one is not too far from threshold, the $\t$ and $\tbar$ quarks
do not have time to hadronize before they decay \cite{toplife}.
The $\b$ and $\bbar$ therefore form part of one colour singlet.
However, because of the time difference between the $\t$ and
$\tbar$ decays, emission of energetic gluons occurs independently
from the $\b$ quark (the $\widehat{\t\b}$ dipole) and from the
$\bbar$ one (the $\widehat{\tbar\bbar}$ dipole). The
$\widehat{\b\bbar}$ dipole is sterile in the same sense as
discussed in section 3.5, i.e. it can only radiate gluons with
$\omega \lesssim \Gamma_{\t}$. Therefore the connection between the
$\b$ and the $\bbar$ (each with accompanying hard gluons) is mainly
provided by the long-distance fragmentation mechanism.

In addition to the observables we have considered, such as
the $\W$ mass, quite a few other interesting phenomena could be
affected as well. Among them are (see also \cite{K6}):
\begin{Enumerate}
\item
The possibility to discriminate between the co-called 1-string and
3-string scenarios \cite{KA} in the process
$\ee \to \t\tbar \to \b\W^+\bbar\W^-$.
\item
The `$\b$-quark attraction effect', namely the possible enhancement
of the amplitude of the $\b$-quarks near the origin in the field of
$\t$-quarks produced in the threshold region of the same process as
above \cite{KB}.
\item
Effects of the $\tbar\b$, $\t\bbar$ and $\b\bbar$ interactions
on the forward--backward asymmetry in $\ee \to \t\tbar$
\cite{KC}.
\item
The azimuthal asymmetry of $\b$-jets in $\t\tbar$ events.
\end{Enumerate}
We therefore expect colour transmutation phenomena to be of
topical interest over the years to come.

\subsection*{Acknowledgements}
This work was supported in part by the UK Science and Engineering
Research Council. The authors are grateful to Yu.L. Dokshitzer and
V.S. Fadin for useful discussions.

\clearpage

\section*{Table}

\begin{table}[h]
\captive{Systematic mass shifts for $\W$ mass determinations at
170 GeV. Results are for the four methods described in the text.
The $\langle \Delta \overmW \rangle$ and $\sigma(\Delta \overmW)$
numbers are the average and spread of the difference between the
reconstructed and the generated $\overmW = (\mW^+ + \mW^-)/2$
mass of an event, defined for the no-reconnection scenario. The
further columns give the additional systematic mass shift obtained
when reconnections are allowed,
$\langle \delta \overmW \rangle =
\langle \Delta \overmW \rangle^{\mrm{recon}} -
\langle \Delta \overmW \rangle^{\mrm{no-recon}} =
\langle \overmW \rangle^{\mrm{recon}} -
\langle \overmW \rangle^{\mrm{no-recon}}$.
(The results for scenario II with $m_0 = 2$ GeV are relative to
no-reconnection numbers with the same cut-off, not shown.
This makes about 5 MeV difference, which would likely vanish
if fragmentation parameters are properly retuned for the higher
$m_0$ scale.)
All numbers are in MeV. The statistical error on
$\langle \Delta \overmW \rangle$ is about 6 MeV and on each
$\langle \delta \overmW \rangle$ about 10 MeV.}

\vspace{5mm}
\begin{center}
\begin{tabular}{|c|c|c|c|c|c|c|c|c|} \hline
Method & $\langle \Delta \overmW \rangle$ & $\sigma(\Delta \overmW)$
& \multicolumn{6}{|c|}{$\langle \delta \overmW \rangle$ (MeV)} \\
\cline{4-9}
  & (MeV) & (MeV) & I & II & II & II$'$ & Inter-  & Instan- \\
  &     &     &  &  & ($m_0=2$) &       & mediate & taneous \\
\hline
1 & $-276$ & 1570 & $+14$ & $-21$ & $-23$ & $-17$ & $+136$ & $+576$ \\
2 & $+158$ & 2060 & $-10$ & $-34$ & $-33$ & $-28$ & $+62$  & $+286$ \\
3 & $-355$ & 2050 & $+14$ & $-34$ & $-31$ & $-27$ & $+88$  & $+496$ \\
4 & $+471$ & 2210 & $+1$  & $-27$ & $-28$ & $-31$ & $+74$  & $+275$ \\
\hline
\end{tabular}

\end{center}
\end{table}

\clearpage

\section*{Figure Captions}

\begin{list}{}{\setlength{\leftmargin}{2cm}
  \setlength{\labelwidth}{1.3cm}\setlength{\labelsep}{0.7cm}
  \setlength{\rightmargin}{0cm}}

\item[Fig. ~1]
Schematic pictures of the toy alternatives of Section 2 (see text).
In (b) to (d) the `+' sign separates (colour singlet) parton
systems that fragment independently of each other.\\
a) Original parton configuration.\\
b) No reconnection.\\
c) Instantaneous reconnection, with reduced showering activity.\\
d) Intermediate reconnection, with same showering as in the
no-reconnection case, but different fragmentation.

\item[Fig. ~2]
Event properties when the two $\W$'s are at rest with a relative
decay angle of $30^{\circ}$, see text.\\
a) Charged multiplicity.\\
b) Rapidity distribution of charged particles.\\
c) Charged multiplicity in the rapidity range $|y| < 1$.\\
d) Charged multiplicity flow in the first quadrant of the
event planes (the other quadrants may be obtained by symmetry).\\
Full histograms for no reconnection, dashed for intermediate, and
dash-dotted for instantaneous reconnection.

\item[Fig. ~3]
Event properties at 170 GeV, for a mixture of $\W^+\W^-$ events.\\
a) Charged multiplicity.\\
b) Rapidity distribution of charged particles.\\
c) Charged multiplicity in the rapidity range $|y| < 1$.\\
d) Difference between reconstructed and generated average
$\W$ mass of an event.\\
Full histograms for no reconnection, dashed for intermediate, and
dash-dotted for instantaneous reconnection.

\item[Fig. ~4]
Diagrams for single primary gluon emission in
$\ee \to \W^+\W^- \to \q_1\qbar_2\q_3\qbar_4$. Here and in what
follows the `radiative blobs' are intended to reflect that the
emission is caused by the conserved-colour quark currents.

\item[Fig. ~5]
An example of a single-gluon decay--decay interference contribution
to the cross section for
$\ee \to \W^+\W^- \to \q_1\qbar_2\q_3\qbar_4$. The gluons are real
or virtual depending on the position of the vertical dashed
line.

\item[Fig. ~6]
Diagrams for double primary gluon emission in
$\ee \to \W^+\W^- \to \q_1\qbar_2\q_3\qbar_4$.

\item[Fig. ~7]
Diagrams for radiatively corrected single gluon emission in
$\ee \to \W^+\W^- \to \q_1\qbar_2\q_3\qbar_4$. Colour rearrangement
could arise as a result of interference between the diagrams of
Fig. 4 and Fig. 7.

\item[Fig. ~8]
An example of the double-gluon decay--decay interference
contribution to the cross section for
$\ee \to \W^+\W^- \to \q_1\qbar_2\q_3\qbar_4$.
Each gluon could be real or virtual depending on the position
of the vertical dashed line.

\item[Fig. ~9]
A $\W$ decaying to $\q\g_1\g_2\g_3\qbar$, with a schematic
representation of the motion of the partons (arrows) and the strings
(thick lines) drawn out between the partons.

\item[Fig. 10]
a) Average kinematical variables in $\W^+\W^-$ events as a function
of c.m. energy. Full the common momentum $p^*$, dashed (dash-dotted)
the minimum (maximum) of the $\W^+$ and $\W^-$ masses.
For comparison, the dotted curve gives the na\"{\i}ve behaviour
$p^* = \sqrt{(E_{cm}/2)^2 - \mW^2}$.  \\
b) Average proper time of $\W^{\pm}$ decays (full curve), and
time (dashed) and space (dash-dotted) separation (in fm) between
the $\W^+$ and $\W^-$ decays, as a function of c.m. energy.

\item[Fig. 11]
Comparison of string topology with the lifetime of the parton taken
into account (a) and not (b). See text for details. Dashed lines show
past motion of partons, arrows indicate direction vectors of motion,
and thick full lines represent the string.

\item[Fig. 12]
${\cal I}(\beta)/{\cal I}(0)$, i.e.\ the overlap of two spherical
colour sources normalized to the value at threshold. Full curve is
for the full expression of eq. (\ref{OOmega}), dashed when
$\W^{\pm}$ are assumed to decay instantaneously,
${\cal P}(\tau^{\pm}) = \delta(\tau^{\pm})$, and dash-dotted
when $\W^{\pm}$ are assumed to decay instantaneously and
additionally the time retardation factor $\theta(t-|\boldx|)$
in eq. (\ref{Omegazero}) is neglected.

\item[Fig. 13]
Probability for colour reconnection, ${\cal P}_{\mrm{recon}}$
according to eq. (\ref{reconsphere}); full curve for $k_0=1$, dashed
for $k_0=5$, and dash-dotted for $k_0=0.2$.

\item[Fig. 14]
Schematic illustration of string topologies, where any gluon has
been represented by either a $\q$ or a $\qbar$ in such a way as to
maintain the correct colour topology. Blobs represent positions of
$\W$ decays, thin dashed lines and arrows indicate motion of string
endpoints, thick dashed and full lines show strings before and after
the reconnection. Figures (a), (b) and (c) are for different
topologies, see text.

\item[Fig. 15]
Probability that a reconnection will occur as a function of (a)
c.m. energy $E_{\mrm{cm}}$ and (b) parton shower cut-off $m_0$.
Full lines for scenario II and dashed ones for scenario I.

\item[Fig. 16]
Probability that a reconnection will occur as a function of (a)
the number of jets in the event and (b) the charged multiplicity.
Full lines for scenario II and dashed ones for scenario I. Kindly
note that statistical fluctuations are not quite negligible in the
few first and last bins.

\item[Fig. 17]
Event properties at 170 GeV, for a mixture of $\W^+\W^-$ events.\\
a) Charged multiplicity.\\
b) Particle-multiplicity distribution around jet axis; see text
for details.\\
Full lines for scenario II, dashed ones for scenario I, dash-dotted
ones for no-reconnection and (in (b)) dotted ones for the
instantaneous scenario.

\item[Fig. 18]
a) Distribution in $\overline{d}_{\mrm{rec}}$, the position of
string reconnection, for 170 GeV $\W^+\W^-$ events. \\
b) Average change in string length,
$\langle \Delta \lambda \rangle$, as a function of
$\overline{d}_{\mrm{rec}}$. \\
Full lines for scenario II, dashed ones for scenario I, and
dotted ones for scenario~II$'$.

\item[Fig. 19]
Difference between the reconstructed and generated average $\W$ mass
of an event. Figures (a)--(d) correspond to methods 1--4 of
Section 5.3 for picking the right pairing of jets. Full histogram
for scenario II, dashed one for scenario I, dash-dotted for
no-reconnection and dotted for the instantaneous scenario.

\end{list}

\end{document}